# Regularizing the pulsar timing array likelihood: A path toward Fourier space

Serena Valtolina[*] and Rutger van Haasteren[†]
*Max Planck Institute for Gravitational Physics (Albert Einstein Institute), Leibniz Universität Hannover, Callinstrasse 38, D-30167, Hannover, Germany*

The recent announcement of evidence for a stochastic background of gravitational waves (GWB) in pulsar timing array (PTA) data has piqued interest across the scientific community. A combined analysis of all currently available data holds the promise of confirming the announced evidence as a solid detection of a GWB. However, the complexity of individual pulsar noise models and the variety of modeling tools used for different types of pulsars present significant challenges for a truly unified analysis. In this work we propose a novel approach to the analysis of PTA data: first a posterior distribution over Fourier modes is produced for each pulsar individually. Then, in a global analysis of all pulsars, these posterior distributions can be reused for a GWB search, which retains all information regarding the signals of interest without the added complexity of the underlying noise models or implementation differences. This approach facilitates combining radio and gamma-ray pulsar data, while reducing the complexity of the model and of its implementations when carrying out a GWB search with PTA data.

## I. INTRODUCTION

Pulsar timing arrays (PTAs) are sensitive to gravitational wave backgrounds (GWBs) in the nano-Hertz frequency band ($10^{-9}$–$10^{-7}$ Hz), a regime where the dominant expected signal is generally expected to arise from supermassive black hole binaries [1–5]. Pulsars are extremely stable, rapidly rotating neutron stars characterized by narrow beams of radio emission. As the pulsar rotates, this collimated emission is detected by a radio telescope as a pulselike signal. By stacking pulsar observations, one obtains high signal-to-noise pulse profiles from which the time of arrival (TOA) for each pulse can be determined accurately using the timing model [see, e.g., [6] ]. The differences between the observed TOAs and the TOAs predicted by the timing model are called timing residuals and can be explained as a combination of different effects, like for example, instrumental noise, low-frequency time-correlated noise due to rotational irregularities, and gravitational wave (GW) induced delays. The dominant GW signal that we expect to observe in the PTA frequency band is a stochastic GWB. This signal affects the TOAs as a low-frequency time-correlated signal present in all pulsar observations. For an unpolarized isotropic GWB, the spatial correlations between the GW-induced residuals of different pulsars follow a quadrupolar pattern, known as the Hellings and Downs (HD) correlation [7]. The HD correlation for a pair of pulsars depends only on the angular separation between them (as viewed from Earth) and predicts a positive correlation when the lines of sight to the pulsars are aligned (and antialigned), and negative correlations when the two lines of sight are almost orthogonal. This is a direct consequence of general relativity and the quadrupolar description of GWs. Detecting the HD correlation is important in a GWB search with PTAs, because this spatial correlation is the feature that allows us to distinguish between GW signal and the various noise processes that contribute to the pulsar timing data.

In the past couple of years, the PTA collaborations [the Chinese PTA (CPTA), the European PTA (EPTA) together with the Indian PTA (InPTA), NANOGrav, the Parkes PTA (PPTA) and the MeerKAT PTA collaborations] have all released their new (radio) datasets and reported evidence for a common red process in their data that shows correlation properties between residuals of different pulsars, consistent with the sought HD correlation [8–12]. A larger number of pulsars and a longer observation time span will increase the sensitivity of the PTA experiments, which we expect to reach the detection threshold within the next few years [13,14].

[*]Contact author: serena.valtolina@aei.mpg.de
[†]Contact author: rutger@vhaasteren.com

The main hypothesis for the source of this GWB signal is the incoherent superposition of continuous GW emission from a population of supermassive black hole binaries [1–5]. Nonetheless, a nano-Hertz frequency signal could be due to GWs generated by early Universe phenomena, such as cosmic strings interactions [e.g., [15,16] ], curvature perturbations [e.g., [17,18] ], quantum chromodynamics (QCD) phase transitions [e.g., [19,20] ], nonstandard inflationary scenarios [e.g., [21–23] ], and more. Those scenarios are comprehensively investigated in the new physics in the early Universe studies of the EPTA + InPTA and NANOGrav collaborations [24,25].

In 2022, the Fermi Large Area Telescope (Fermi-LAT) collaboration published the first PTA analysis of gamma-ray pulsars [26], complementing radio PTAs despite unique observational challenges. The Fermi-LAT, a space-based observatory sensitive to 20 MeV–300 GeV photons, has detected over 200 gamma-ray pulsars, enabling GW searches analogous to radio PTAs. However, gamma-ray timing differs fundamentally from radio methods. In radio observations, pulses are folded into high signal-to-noise profiles by averaging over the spin period [6], allowing precise TOA determination against a telescope's atomic clock. For Fermi-LAT, individual gamma-ray arrival times and energies are recorded, with each photon weighted by its probability of originating from the pulsar versus background [27–29]. This uncertainty—combined with low fluxes—often necessitates prohibitively long integration times for traditional folding. To address this, [30] proposed a *photon-to-photon* approach, bypassing folded profiles entirely. Updates on the second gamma-ray PTA data release are detailed in [31].

While the basic processing of TOAs and photons for both gamma-ray pulsars and radio pulsars can be done with PINT and TEMPO2—software packages for pulsar timing [32–35]—the search for gravitational waves requires more specific modeling techniques that are usually carried out using specialized software implementations. Historically, the fundamental differences between the radio data and the gamma-ray data have caused implementations of GW search to be highly specialized: packages like ENTERPRISE [36] or FORTY-TWO [37] are designed specifically for radio TOAs, whereas the method of [30] is only able to analyze gamma-ray photon data. Modifying these packages for a combined radio and gamma-ray dataset is highly nontrivial, and requires significant development work. Our work makes this joint analysis easier.

PTA data is typically analyzed using Bayesian inference, where the posterior distribution proportional to a likelihood function times a prior distribution is explored to obtain the best estimate of the model parameters. The likelihood function used for inference on radio PTA data [time-domain likelihood, [38–41] ] is very general and flexible, allowing the inclusion of many independent processes to be modeled simultaneously, including processes that cause correlations between timing residuals of different pulsars. For gamma-ray data, instead, applications of the photon-to-photon approach have only produced upper limits on the amplitude of a possible GWB signal [31]. In this approach correlation information between pulsar pairs is not used, and instead the model contains a process that is assumed to be common but uncorrelated among all pulsars. A complication arises when trying to combine the radio and gamma-ray data in a single analysis: the likelihood functions used for the two types of data are not implemented in a single analysis code. Ideally, we would preprocess the data in a way that allows us to write the likelihood function in a common form, so that we can analyze the data jointly. The difficulty with that approach is that the likelihood function, when combined with an improper prior on certain parameters $\mathbf{b}$ (defined later), is not normalizable with respect to $\mathbf{b}$. In our paper, we present a regularization of the likelihood function that allows us to write it as a Gaussian distribution with respect to the model parameters. This process moves the analysis to the Fourier domain and allows the inclusion of correlated signals in the model (see Appendix A for a detailed discussion on the meaning of "regularization"). Most importantly, this method can be applied to both gamma-ray and radio data, independently of the package used to interpret the timing data and build the signal model.

The other key advantage of using the method introduced in this paper is that it allows us to divide the GWB search into two steps. The first step focuses on individual pulsars and carries out inference on signals that are not covariant with a GWB and other low-frequency time-correlated effects. In the second step the analysis is carried out on a full array of pulsars, where the focus lies on the time-correlated stochastic signals, including the GWB.

This paper shows the analytical derivation of a regularized formulation for the PTA likelihood in the Fourier domain, and presents some results from inference on real data. In detail, in Sec. II, we present a quick review of the time-domain likelihood definition (Sec. II B) and then derive in detail the regularized formulation in the Fourier domain (Sec. II C). Section III presents results for Bayesian inference runs on the EPTA `DR2new` dataset [42]. In particular, we show the comparison between posteriors obtained with the two likelihood formulations for the case of a single pulsar noise analysis (SPNA) on J1738+0333 (Sec. III A), and the results of a GWB search on the whole pulsars array (Sec. III B). We also briefly discuss applying this algorithm to the gamma-ray PTA dataset in Sec. III C. We conclude by discussing advantages and future directions in Sec. IV.

## II. METHODS: TIME VS FOURIER DOMAIN

The primary data of a PTA analysis consists of the set of TOAs for an array of pulsars. The TOAs can be rewritten as the sum of a deterministic arrival time $f(t)$, which is well-modeled by the timing model (including relativistic and propagation effects together with spin, spin-down, binary orbit modeling, etc.), and stochastic delays. We model stochastic delays as zero-mean time-correlated random processes for which we only parametrize the (auto)correlation function. The timing model definition and the evaluation of the expected TOAs at the Solar System barycenter is done using pulsar timing software packages like TEMPO2 [34,35] and PINT [32,33]. The stochastic delays modeling includes signals such as measurement errors, pulsar low-frequency noise induced by rotational instabilities, dispersion due to the interstellar medium, and GW signals. In general, the observed times of arrival can be written as

$$\begin{aligned} T^{\mathrm{obs}} &= f(t;\boldsymbol{\beta}) + \delta\mathbf{t} \\ &= f(t;\boldsymbol{\beta}) + \delta\mathbf{t}_{\mathrm{WN}} + \delta\mathbf{t}_{\mathrm{RN}} + \delta\mathbf{t}_{\mathrm{DM}} + \delta\mathbf{t}_{\mathrm{GW}} + \cdots \end{aligned} \quad (1)$$

where $\delta\mathbf{t} \equiv \sum_j \delta\mathbf{t}_{(\mathrm{j})}$ are the various stochastic signal delays and $f(t;\boldsymbol{\beta})$ are the TOAs predicted by the timing model for the model parameters $\boldsymbol{\beta}$. The term $\delta\mathbf{t}_{\mathrm{WN}}$ refers to measurements errors and other processes uncorrelated in time (white noise), $\delta\mathbf{t}_{\mathrm{RN}}$ to pulsar spin noise, frequency-dependent delays due to interaction of the pulses with the interstellar medium (DM variations) are modeled by $\delta\mathbf{t}_{\mathrm{DM}}$, and $\delta\mathbf{t}_{\mathrm{GW}}$ refers to the GW-induced delays. In Sec. II A, we show how each term of Eq. (1) is modeled.

In PTA data analysis, Bayesian inference is one of the most common strategies for obtaining information about model parameters. Given a data set $D$, the posterior probability distribution $p(\Theta|D)$ for each parameter $\Theta$ of the model is proportional to the product between a likelihood $p(D|\Theta)$ and a prior function $p(\Theta)$. The likelihood function is defined as the probability density function of the data conditioned on the model and the model parameters. In Sec. II B, we present a brief overview of the PTA likelihood as it is coded in ENTERPRISE, which is how it is often used inference on real data from PTA collaborations. The main references for this section are [36,38–41]. In Sec. II C, instead, we present our alternative formulation of the PTA likelihood in Fourier domain.

## A. TOAs: Signal model components

We give a brief overview of the various signal model components.

### 1. Timing model

The timing residuals $\delta\mathbf{t}$ are obtained from the observed $T^{\mathrm{obs}}$ as $\delta\mathbf{t} = T^{\mathrm{obs}} - f(t;\boldsymbol{\beta}_0)$ [Eq. (1)], where the term $f(t;\boldsymbol{\beta}_0)$ corresponds to the TOAs predicted by the timing model evaluated at the reference values $\boldsymbol{\beta}_0$ (usually best-guesses from the timing analysis) for the timing parameters $\boldsymbol{\beta}$. From Eq. (1), we see that $\delta\mathbf{t}$ can be rewritten as a sum of stochastic delay components. We now describe in detail the model for each of those components.

The time-delays due to timing model ephemeris offsets $\delta\mathbf{t}_{\mathrm{TM}}$ are defined as the first-order linearization around the reference parameters $\boldsymbol{\beta}_0$

$$f(t;\boldsymbol{\beta}) = f(t;\boldsymbol{\beta}_0) + \delta\mathbf{t}_{\mathrm{TM}} = f(t;\boldsymbol{\beta}_0) + M\boldsymbol{\xi} + \mathcal{O}(\boldsymbol{\xi}^2). \quad (2)$$

The matrix $M$ is called the design matrix and, given a timing model $f(t;\boldsymbol{\beta})$, it is defined as the matrix of the partial derivatives of the timing model with respect to the timing model parameters: $M_{jk} \equiv (\partial f(t_j;\boldsymbol{\beta})/\partial\boldsymbol{\beta}_k)|_{\boldsymbol{\beta}_0}$. The vector $\boldsymbol{\xi}$ represents the ephemeris offsets: $\boldsymbol{\xi} \equiv \boldsymbol{\beta} - \boldsymbol{\beta}_0$.

### 2. White noise

The TOAs are calculated from averaged pulse profiles by comparing the observed pulse profile with the template profile. From this comparison, we can obtain the TOAs with a certain measurement error $\sigma_{\mathrm{TOA}}$. If this process were perfect, the root-mean-square of $\delta\mathbf{t}_{\mathrm{WN}}$ would be $\sigma_{\mathrm{TOA}}$. Due to inaccuracies in this process, and other potential instrumental effects, the estimated measurement error $\sigma_{\mathrm{TOA}}$ needs to be modified or scaled to properly account for the observed variance in $\delta\mathbf{t}_{\mathrm{WN}}$. These modeling "calibration parameters," while somewhat motivated, need to be determined from the data. The usual way to do this is with two parameters, called [EFAC ($\mathcal{E}$) and EQUAD ($\mathcal{Q}$)], which are specific for each observing system (particular configuration of observing backend and receiver)

$$\langle \delta\mathbf{t}_{\mathrm{WN},i} \delta\mathbf{t}^T_{\mathrm{WN},j} \rangle_{\mathrm{pr}} = \mathcal{E}^2 \sigma^2_{\mathrm{TOA},i}\, \delta_{ij} + \mathcal{Q}^2 \delta_{ij}, \quad (3)$$

where the indices $i$ and $j$ label the observation.[1] An additional white noise component describing the pulse phase jitter [commonly known as jitter noise or ECORR ($\mathcal{J}$)] can also be included. This additional noise parameter models correlated white noise between TOAs observed at the same epoch at different radio frequencies. It can be added to Eq. (3) as an additional parameter similar to EQUAD: $\mathcal{J}^2\delta_{ef}$, where $e$ and $f$ label the observed epochs. Thus, for ECORR parameters, different observations at the same observing epoch are correlated with covariance $\mathcal{J}^2$.

### 3. Red noise and chromatic noise

The turbulent ionized interstellar medium (ISM), the solar wind, and similar effects all influence the pulse propagation differently at different wavelengths, causing a frequency-dependent delay. This effect is modeled as an observing-frequency dependent time-correlated delay $\delta\mathbf{t}_{\mathrm{DM}}$, which is proportional to the square of the inverse of the observing frequency $\nu_{\mathrm{obs}}$. Scattering by the ISM is sometimes similarly modeled, with the difference that the observation frequency dependence is not a simple squared relationship. We refer to these observing-frequency dependent delays as "chromatic" delays.

[1] In this paper, we use $\langle\cdot\rangle$ to indicate the covariance under the posterior distribution: $\langle ab\rangle = \mathrm{cov}(a, b)$. When we evaluate the covariance under the prior distribution instead, we use the notation $\langle\cdot\rangle_{\mathrm{pr}}$.

Pulsars are very stable rotating bodies. However, various effects such as the coupling of the magnetosphere to the pulsar will cause subtle random walk behavior in the rotational frequency of the pulsar. The effect is a time-correlated random process that is often referred to as "timing noise" or "intrinsic red noise." We group all time-correlated effects together, and refer to it as achromatic red noise, where the delays are written as $\delta\mathbf{t}_{\mathrm{RN}}$.

The chromatic ($\delta\mathbf{t}_{\mathrm{DM}}$) and achromatic ($\delta\mathbf{t}_{\mathrm{RN}}$ and $\delta\mathbf{t}_{\mathrm{GW}}$) low frequency processes are modeled as Gaussian processes [41]. Without loss of generality, they are written as a discrete sum of cosine and sine functions evaluated at specific frequencies

$$\delta\mathbf{t}_{\mathrm{RN/DM}} = \sum_k [a_k \cos(2\pi kt/T) + b_k \sin(2\pi kt/T)]\nu_{\mathrm{obs}}^{\alpha}, \quad (4)$$

where $T$ is the total time of observation, and $\alpha = 0$ for RN and $\alpha = -2$ for DM variatons. In matrix notation

$$\begin{aligned}\delta\mathbf{t}_{\mathrm{RN}} + \delta\mathbf{t}_{\mathrm{GW}} &= F\mathbf{a} \\ \delta\mathbf{t}_{\mathrm{DM}} &= F_{\mathrm{DM}}\mathbf{a}_{\mathrm{DM}},\end{aligned} \quad (5)$$

where $\mathbf{a}$ and $\mathbf{a}_{\mathrm{DM}}$ are the Fourier coefficients of $\delta\mathbf{t}_{\mathrm{RN}}$ and $\delta\mathbf{t}_{\mathrm{DM}}$ respectively. The matrices $F$ and $F_{\mathrm{DM}}$ (often referred to as Fourier design matrices) are the linear transformation corresponding to a discrete Fourier transform with the "backward" normalization [conventions as in SciPy [43] ]. The Fourier design matrix for the chromatic noise has an additional factor that includes the dependence upon the observing radio frequency $\nu_{\mathrm{obs}}$: $F_{\mathrm{DM},ij} = F_{ij} \times (\nu_{\mathrm{obs}} / 1400\ \mathrm{MHz})^{-2}$. In the rest of this paper, we will omit the chromatic delays $\delta\mathbf{t}_{\mathrm{DM}}$ for simplicity and clarity of presentation. In everything that follows, it is straightforward to include the chromatic delays in the analysis by adding the corresponding Fourier coefficients $\mathbf{a}_{\mathrm{DM}}$ and the Fourier design matrix $F_{\mathrm{DM}}$. We refer to the Fourier coefficients as $\mathbf{a}$ and the Fourier design matrix as $F$.

Note that the Fourier modes are a complete basis; thus, even a periodic function is well represented as a sum of Fourier modes within its domain. While this may introduce discontinuities at the edges of the domain of the periodic function, in practice this is compensated by the marginalization over the spin-down parameters [44].

The prior covariance matrix of the Fourier coefficients $\phi = \langle \mathbf{a}\mathbf{a}^T \rangle_{\mathrm{pr}}$ depends upon the hyperparameters $\boldsymbol{\rho}$ of the time-correlated process

$$\begin{aligned}[\phi]_{(a,b)(j,k)} \equiv\ & \Gamma_{ab}\delta_{jk} S_{\mathrm{GWB}}(f_j; \boldsymbol{\rho}_{\mathrm{GWB}}) \\ & + \delta_{ab}\delta_{jk} S_{\mathrm{RN}}(f_j; \boldsymbol{\rho}_{\mathrm{RN}}),\end{aligned} \quad (6)$$

where $a$, $b$ label the pulsar, $j$, $k$ label the frequency components, $\Gamma_{ab}$ is the HD correlation, and $S(f;\boldsymbol{\rho})$ is the power spectral density at frequency $f$, described by the hyperparameters $\boldsymbol{\rho}$. In absence of pulsar-correlated signals, the $\phi$ matrix [defined as in Eq. (6)] is assumed to be diagonal [44]; although, recent work shows how to properly calculate from first principles the true $\phi$, which allows for correlations between different frequency components referred to the same pulsar [45,46]. A $\phi$ defined in this way would be block-diagonal in absence of pulsar correlated signals.

The pulse arrival time delays we just described are usually written with the following compact notation:

$$M\boldsymbol{\xi} + F\mathbf{a} = T\mathbf{b}. \quad (7)$$

See Table I for a complete summary of the notation used through this paper.

### B. The time-domain PTA likelihood

In this section, we present a brief overview of the PTA likelihood as it is coded in ENTERPRISE and used for inference on real data from PTA collaborations. The main references for this section are [36,38–41].

When inferring the noise properties of a single pulsar, the posterior distribution for the model parameters $(\mathbf{b}, \boldsymbol{\rho}, \boldsymbol{\theta})$ (product of likelihood function and prior distribution) is usually written as

$$\begin{aligned} p(\mathbf{b}, \boldsymbol{\rho}, \boldsymbol{\theta} | \delta\mathbf{t}) p(\delta\mathbf{t}) &= p(\delta\mathbf{t} | \mathbf{b}, \boldsymbol{\theta}) p(\mathbf{b} | \boldsymbol{\rho}) p(\boldsymbol{\rho}) p(\boldsymbol{\theta}) \\ p(\delta\mathbf{t} | \mathbf{b}, \boldsymbol{\theta}) &= \frac{\exp\left[-\frac{1}{2}(\delta\mathbf{t} - T\mathbf{b})^T N^{-1} (\delta\mathbf{t} - T\mathbf{b})\right]}{\sqrt{\det(2\pi N)}} \\ &= \mathcal{N}(\delta\mathbf{t} | T\mathbf{b}, N) \\ p(\mathbf{b} | \boldsymbol{\rho}) &= \mathcal{N}(\mathbf{b} | 0, B), \end{aligned} \quad (8)$$

where $N \equiv \langle \delta\mathbf{t}_{\mathrm{WN}} \delta\mathbf{t}_{\mathrm{WN}}^T \rangle_{\mathrm{pr}}$, and $T$ and $\mathbf{b}$ are defined in Eq. (7). We introduce the notation that $\mathcal{N}(\mathbf{x} | \boldsymbol{\mu}, \Sigma)$ represents a multivariate Gaussian distribution in the parameter $\mathbf{x}$ with mean $\boldsymbol{\mu}$ and covariance $\Sigma$. The quantity $P(\delta\mathbf{t})$ is the evidence or fully marginalized likelihood, often denoted as $Z$. Going forward, we omit the $P(\delta\mathbf{t})$ occasionally and instead use a $\propto$ for readability. $\boldsymbol{\theta}$ represents the white noise parameters. The second exponential on the right-hand side (rhs) describes the conditioned probability of $\mathbf{b}$ upon the model hyperparameters $\boldsymbol{\rho}$. The prior matrix $B$ is a block-diagonal matrix defined as

$$B \equiv \begin{bmatrix} \infty & 0 \\ 0 & \phi \end{bmatrix} \tag{9}$$

where we assigned an improper infinite prior to the timing model ephemeris offsets $\boldsymbol{\xi}$. Using improper priors on $\boldsymbol{\xi}$ is customary in pulsar timing. Although adding more realistic Gaussian priors on $\boldsymbol{\xi}$ is trivial [41], the data is so informative with respect to the prior that there is no practical need to change the practice of using improper priors. This is partly because the PTA datasets are constructed in such a way that model parameters that are not well constrained by the data are not included in the model.

TABLE I. Notation summary.

| Symbol | Description |
|---|---|
| $\boldsymbol{\beta}$ | Timing model parameters ($\boldsymbol{\beta}_0$ are the best-fit values) |
| $T^{\rm obs}$ | Observed TOAs |
| $f(t;\boldsymbol{\beta})$ | Timing model predicted TOAs for the model parameters $\boldsymbol{\beta}$ |
| $\delta\mathbf{t}$ | Timing residuals $\delta\mathbf{t} \equiv T^{\rm obs} - f(t;\boldsymbol{\beta}_0)$ |
| $M$ | Design matrix $M_{jk} \equiv (\partial f(t_j;\boldsymbol{\beta})/\partial\boldsymbol{\beta}_k)\|_{\boldsymbol{\beta}_0}$ |
| $\boldsymbol{\xi}$ | Ephemeris offsets $\boldsymbol{\xi} \equiv \boldsymbol{\beta} - \boldsymbol{\beta}_0$ |
| $\sigma_{\rm TOA}$ | Measurement error associated to the TOA |
| $F$ | Fourier design matrix |
| $\mathbf{a}$ | Fourier coefficients |
| $T$ | $T \equiv [M, F]$ |
| $\mathbf{b}$ | $\mathbf{b} \equiv \begin{bmatrix} \boldsymbol{\xi} \\ \mathbf{a} \end{bmatrix}$ |
| $\boldsymbol{\theta}$ | Deterministic signals and white noise parameters |
| $\boldsymbol{\rho}$ | Noise hyperparameters |
| $\phi$ | $\phi \equiv \langle \mathbf{a}\mathbf{a}^T \rangle_{\rm pr}$ [Eq. (6)] |
| $B$ | Prior matrix $B \equiv {\rm diag}(\infty, \phi)$, $B^{-1} = {\rm diag}(0, \phi^{-1})$ |
| $N$ | White noise covariance matrix $N \equiv \langle \delta\mathbf{t}_{\rm WN}\delta\mathbf{t}_{\rm WN}^T \rangle_{\rm pr}$ |
| $N_p$ | Number of pulsars |
| $C$ | Covariance matrix of the fully marginalized likelihood: $C \equiv N + TBT^T$ |
| $\tilde{N}$ | $\tilde{N}^{-1} \equiv N^{-1} - N^{-1}M(M^T N^{-1} M)^{-1} M^T N^{-1}$. |
| $\Sigma$ | $\Sigma^{-1} \equiv F^T \tilde{N}^{-1} F + \phi^{-1}$ |
| $\hat{\mathbf{a}}$ | Optimal estimator of the Fourier coefficients $\hat{\mathbf{a}} \equiv \Sigma F^T \tilde{N}^{-1} \delta\mathbf{t}$ |
| $\hat{\mathbf{a}}_0$ | Mean of the normal distribution $p(\mathbf{a}\|\delta\mathbf{t}, \boldsymbol{\rho}_0)$ [Eq. (19)] |
| $\Sigma_0$ | Variance of the normal distribution $p(\mathbf{a}\|\delta\mathbf{t}, \boldsymbol{\rho}_0)$ [Eq. (19)] |
| $\mathcal{N}(\mathbf{x}\|\boldsymbol{\mu}, \Sigma)$ | Multivariate Gaussian distribution in the parameter $\mathbf{x}$ with mean $\boldsymbol{\mu}$ and covariance $\Sigma$ |

The posterior distribution in Eq. (8) can be generalized to the case of an array of pulsars

$$p(\mathbf{b}, \boldsymbol{\rho}, \boldsymbol{\theta}|\delta\mathbf{t})p(\delta\mathbf{t}) = \left[\prod_{k=1}^{N_p} p(\delta\mathbf{t}_k|\mathbf{b}_k, \boldsymbol{\theta})\right] p(\mathbf{b}|\boldsymbol{\rho})p(\boldsymbol{\rho})p(\boldsymbol{\theta}), \tag{10}$$

where $N_p$ is the number of pulsars and the terms $\mathbf{b}$ and $\delta\mathbf{t}$ refer (from here onward) to the concatenation of the elements $\mathbf{b}_k$ and $\delta\mathbf{t}_k$ for all pulsars.

Carrying out a Bayesian inference run with the full likelihood in Eq. (10) would be challenging to sample because of the very high number of parameters, combined with extreme curvature of the posterior distribution [47]. Usually, the analysis aims to obtain estimates of the hyperparameters $\boldsymbol{\rho}$. To do that, we use the posterior distribution marginalized over the timing model parameters and Fourier coefficients. Integrating Eq. (10) over d$\mathbf{b}$ we obtain

$$\begin{aligned} p(\boldsymbol{\rho}, \boldsymbol{\theta}|\delta\mathbf{t})p(\delta\mathbf{t}) &= \int {\rm d}\mathbf{b} \left[\prod_{k=1}^{N_p} p(\delta\mathbf{t}_k|\mathbf{b}_k, \boldsymbol{\theta})\right] p(\mathbf{b}|\boldsymbol{\rho})p(\boldsymbol{\rho}, \boldsymbol{\theta}) \\ &= \mathcal{N}(\delta\mathbf{t}|0, C)p(\boldsymbol{\rho}, \boldsymbol{\theta}), \end{aligned} \tag{11}$$

where $C \equiv N + TBT^T$ and we combined the priors $p(\boldsymbol{\rho}, \boldsymbol{\theta}) = p(\boldsymbol{\rho})p(\boldsymbol{\theta})$. See [38,41] for more detailed descriptions of the PTA likelihood marginalization.

### C. The regularized PTA likelihood

The main contribution of this work concerns the likelihood $P(\delta\mathbf{t}|\mathbf{b}, \boldsymbol{\theta})$, as defined in Eq. (8)

$$\begin{aligned} p(\delta\mathbf{t}|\mathbf{b}, \boldsymbol{\theta}) &= \frac{\exp\left[-\frac{1}{2}(\delta\mathbf{t} - T\mathbf{b})^T N^{-1}(\delta\mathbf{t} - T\mathbf{b})\right]}{\sqrt{\det(2\pi N)}} \\ &= \mathcal{N}(\delta\mathbf{t}|T\mathbf{b}, N), \end{aligned} \tag{12}$$

which we would like to approximate using a multivariate Gaussian distribution in the parameters $\mathbf{b}$. This seems possible on the surface, because the $\mathbf{b}$ enter quadratically in the exponential. However, the problem is that the likelihood function is not normalizable in this form. The reason for this is that the column space of $T$ may contain linear dependencies, which means that the rank of $T$ is smaller than the number of columns. In other words, some (combination of) Fourier modes can be written as a linear combination of timing model parameters. This is strictly a consequence of combining the set of basis vectors of $M$ and $F$, because during the process of creating PTA data releases, care is taken to remove such dependencies within the timing model. The matrix $F$ represents an analog of a discrete Fourier transform, which means that the columns are also not linearly dependent either.

The above problem can be solved by reducing the freedom of the model parameters $\mathbf{b}$ using a restriction, or *regularization*, in the model. This is common practice in the field of machine learning and statistics when a model would otherwise overfit the data, and solutions for this problem typically involve penalty factors, priors, or other types of model restrictions. Ridge regression [48] is a well-known example of this, where a penalty term (Gaussian prior) is added to the model to prevent overfitting and parameter degeneracies. The formulation of ridge regression as a Gaussian prior is especially compatible with our Gaussian process modeling of the timing residuals, which makes it the natural choice for our purposes.

The ridge regression regularization is imposed via a Gaussian prior on the Fourier coefficients, where the prior covariance is defined by a set of reference hyperparameters $\boldsymbol{\rho}_0$, that we are free to choose: $p(\mathbf{b}|\boldsymbol{\rho}_0)$. Because both terms are quadratic in $\mathbf{b}$ inside the exponential, we may write

$$p(\delta\mathbf{t}|\mathbf{b},\boldsymbol{\theta})p(\mathbf{b}|\boldsymbol{\rho}_0) \propto \mathcal{N}(\mathbf{b}|\hat{\mathbf{b}},\Sigma_b) \tag{13}$$

where $\hat{\mathbf{b}}$ and $\Sigma_b$ are the mean and covariance of the multivariate Gaussian distribution, both of which depend on the observations $\delta\mathbf{t}$ and the reference parameters $\boldsymbol{\rho}_0$. The mean $\hat{\mathbf{b}}$ is the optimal estimator of the Fourier coefficients $\mathbf{a}$ given the data $\delta\mathbf{t}$ and the reference parameters $\boldsymbol{\rho}_0$, and $\Sigma_b$ is the corresponding covariance

$$\begin{aligned}\Sigma_b^{-1} &= T^T N^{-1} T + B_0^{-1} \\ \hat{\mathbf{b}} &= \Sigma_b T^T N^{-1} \delta\mathbf{t}.\end{aligned} \tag{14}$$

where $B_0$ is the prior matrix defined in Eq. (9) with the prior covariance matrix $\phi$ replaced by the reference covariance matrix $\phi_0$, which is a function of the reference hyperparameters $\boldsymbol{\rho}_0$. Note that the regularized likelihood of Eq. (13) written as a multivariate normal distribution is an identity, and it contains all information of our data regarding signals that can be expressed in terms of the parameters $\mathbf{b}$.

### D. Two-step analysis

The regularized likelihood formulation presented in the previous Section is useful, because it is easily represented in any computing environment. All the complexity of pulsar timing data, whether from gamma-ray data or from radio observations, is compressed in just the mean $\hat{\mathbf{b}}_0$ and covariance $\Sigma_b$ of the multivariate Gaussian distribution.[2] Of course, we placed a dummy prior $p(\mathbf{b}|\boldsymbol{\rho}_0)$ on our likelihood in order to regularize it. Our aim now is to use this representation of the data to carry out inference with our *actual* model, in which the ridge regression prior $p(\mathbf{b}|\boldsymbol{\rho}_0)$ is replaced with $p(\mathbf{b}|\boldsymbol{\rho})$.

So, our full approach can be summarized as a two-step analysis:

*Step 1*: inference on and marginalization over the parameters of all the signal processes that are not covariant with the processes we model in the Fourier domain (like the GWB). Those are white noise and deterministic signals like, for example, DM dips. In this step, each pulsar is analyzed *individually*. The noise signals covariant with a GWB (RN, DM variations, etc.) are still included in the model, but their hyperparameters are fixed $\boldsymbol{\rho} = \boldsymbol{\rho}_0$. The end result is an approximation of the posterior distribution as a multivariate Gaussian distribution in the Fourier coefficients $\mathbf{a}$.

*Step 2*: inference on the hyperparameters of the GWB and all signals that are typically modeled in terms of the Fourier coefficients (RN, DM variations, chromatic noise, and similar processes). Here, the analysis runs over the *whole array* of pulsars.

The first step is carried out on a per-pulsar basis, which means that all pulsars can be processed in parallel. The second step, which is carried out on the whole array, becomes simpler to implement in practice. We found it straightforward to do without the need for contemporary pulsar timing analysis packages like ENTERPRISE. Due to the advanced caching mechanisms in ENTERPRISE, the second step is not faster in our approach than a typical run with ENTERPRISE.

#### 1. Step 1: Per-pulsar analysis

In *Step 1*, we analyze each pulsar individually with a ridge regression regularization prior. Let us start with the marginalization over $\boldsymbol{\xi}$ only

[2] Note that in this section we will first marginalize over $\boldsymbol{\xi}$, so the quantities $\mathbf{b}$, $T$ will be replaced by $\mathbf{a}$, $F$.

$$
\begin{aligned}
p(\delta\mathbf{t}|\mathbf{a},\boldsymbol{\theta}) &= \int \mathrm{d}\boldsymbol{\xi} P(\delta\mathbf{t}|\mathbf{b},\boldsymbol{\theta}) \\
&= \int \mathrm{d}\boldsymbol{\xi}\mathcal{N}(\delta\mathbf{t}|M\boldsymbol{\xi}+F\mathbf{a},N) \\
&= \frac{\exp\left[-\frac{1}{2}(\delta\mathbf{t}-F\mathbf{a})^T\tilde{N}^{-1}(\delta\mathbf{t}-F\mathbf{a})\right]}{\sqrt{\det(2\pi N)}\sqrt{\det(2\pi M^T N^{-1}M)}}, \quad (15)
\end{aligned}
$$

where $\tilde{N}^{-1} \equiv N^{-1} - N^{-1}M(M^T N^{-1}M)^{-1}M^T N^{-1}$. It is tempting to write Eq. (15) as $\mathcal{N}(\delta\mathbf{t}|F\mathbf{a},\tilde{N})$, but that would not be formally correct: $\tilde{N}$ is not a valid covariance matrix as it is not positive definite.

Next, we include our ridge regression prior $P(\mathbf{a}|\boldsymbol{\rho}_0)$ to regularize the distribution as a function of $\mathbf{a}$, and write it as

$$
p(\delta\mathbf{t}|\mathbf{a},\boldsymbol{\theta})p(\mathbf{a}|\boldsymbol{\rho}_0) \propto \mathcal{N}(\mathbf{a}|\hat{\mathbf{a}}'_0,\Sigma'_0), \quad (16)
$$

where we define $\hat{\mathbf{a}}'_0$ and $\Sigma'_0$ as

$$
\begin{aligned}
\Sigma_0'^{-1} &= F^T\tilde{N}^{-1}F + \phi_0^{-1} \\
\hat{\mathbf{a}}'_0 &= \Sigma'_0 F^T \tilde{N}^{-1}\delta\mathbf{t}. \quad (17)
\end{aligned}
$$

Keep in mind here that $N$, $\tilde{N}$, $\Sigma'_0$, and $\hat{\mathbf{a}}'_0$ are functions of the white noise parameters $\boldsymbol{\theta}$. So far everything is an analytical identity. Next, we also marginalize over the parameters $\boldsymbol{\theta}$

$$
\begin{aligned}
&\int \mathrm{d}\boldsymbol{\theta} p(\delta\mathbf{t}|\mathbf{a},\boldsymbol{\theta})p(\mathbf{a}|\boldsymbol{\rho}_0)p(\boldsymbol{\theta}) \\
&\propto \int \mathrm{d}\boldsymbol{\theta}\mathcal{N}(\mathbf{a}|\hat{\mathbf{a}}'_0,\Sigma'_0)p(\boldsymbol{\theta}) \\
&\quad \times \int \mathrm{d}\boldsymbol{\theta}\mathcal{N}(\mathbf{a}|\hat{\mathbf{a}}'_0,\Sigma'_0)p(\boldsymbol{\theta}) \approx \mathcal{N}(\mathbf{a}|\hat{\mathbf{a}}_0,\Sigma_0). \quad (18)
\end{aligned}
$$

That particular analysis is the analog of the single pulsar noise analysis (SPNA), but now with fixed hyperparameters $\boldsymbol{\rho}=\boldsymbol{\rho}_0$ acting as our regularization prior. The only constraints in choosing the values $\boldsymbol{\rho}_0$ are given by the numerical stability of the method. In brief, we want our ridge regression prior $p(\mathbf{a}|\boldsymbol{\rho}_0)$ to be small enough do that, in the renormalization step, the covariance matrices involved are positive definite (more details can be found in Appendix A). The approximation in the second line is quite accurate when the parameters $\boldsymbol{\theta}$ are not strongly correlated with the Fourier coefficients $\mathbf{a}$, which is the case for the white noise parameters and various kinds of other processes that are not covariant with the GWB, such as, for example, DM dips. All processes covariant with the GWB need to be represented in Fourier domain and included in *Step 2*. The mean $\hat{\mathbf{a}}_0$ and covariance $\Sigma_0$ of the multivariate Gaussian need to be determined numerically. We assume the analysis has been done with a Markov chain Monte Carlo (MCMC) method, which is the most common approach in PTA data analysis. At the $i$th iteration of the MCMC, we have a the noise parameters $\boldsymbol{\theta}_i$. Using Eq. (17) we can then compute the mean $\hat{\mathbf{a}}'_{0,i}$ and covariance $\Sigma'_{0,i}$ for each sample. From that, if we assume the total distribution of $\mathbf{a}$ is the Gaussian $\mathcal{N}(\mathbf{a}|\hat{\mathbf{a}}_0,\Sigma_0)$, we obtain

$$
\begin{aligned}
\hat{\mathbf{a}}_0 &= \mathrm{mean}(\hat{\mathbf{a}}_{0,i}) \\
\Sigma_0 &= \frac{1}{N_s-1}\left[\sum_{i=1}^{N_s}(\Sigma_{0,i}+\hat{\mathbf{a}}_{0,i}\hat{\mathbf{a}}_{0,i}^T) - N_s\hat{\mathbf{a}}_0\hat{\mathbf{a}}_0^T\right], \quad (19)
\end{aligned}
$$

where $N_s$ is the number of samples considered from the white noise chain. This method to approximate the mean $\hat{\mathbf{a}}_0$ and variance $\Sigma_0$ of the normal distribution would correspond to reconstructing the distribution from a large number of Fourier coefficients samples. See Appendix B for a complete derivation of Eq. (19).

### 2. Step 2: Full array analysis

The analysis of *Step 1* was carried out with a regularization prior where $\boldsymbol{\rho}=\boldsymbol{\rho}_0$ was held fixed. In reality, we are actually interested in $p(\mathbf{a},\boldsymbol{\rho}|\delta\mathbf{t})$. Following Eq. (18), we can write

$$
\begin{aligned}
p(\delta\mathbf{t},\mathbf{a},\boldsymbol{\rho}) &= \int \mathrm{d}\boldsymbol{\theta} p(\delta\mathbf{t}|\mathbf{a},\boldsymbol{\theta})p(\mathbf{a}|\boldsymbol{\rho})p(\boldsymbol{\theta})p(\boldsymbol{\rho}) \\
&\approx \int \mathrm{d}\boldsymbol{\theta} p(\delta\mathbf{t}|\mathbf{a},\boldsymbol{\theta})p(\mathbf{a}|\boldsymbol{\rho}_0)p(\boldsymbol{\theta})p(\boldsymbol{\rho})\frac{p(\mathbf{a}|\boldsymbol{\rho})}{p(\mathbf{a}|\boldsymbol{\rho}_0)} \\
p(\mathbf{a},\boldsymbol{\rho}|\delta\mathbf{t}) &\approx \mathcal{N}(\mathbf{a}|\hat{\mathbf{a}}_0,\Sigma_0)\frac{p(\mathbf{a}|\boldsymbol{\rho})}{p(\mathbf{a}|\boldsymbol{\rho}_0)}p(\boldsymbol{\rho}). \quad (20)
\end{aligned}
$$

We used Eq. (18) on the last line, and $p(\boldsymbol{\rho})$ is the prior on the hyperparameters. We see that we can get the full-array posterior distribution from the per-pulsar posterior distribution approximation simply by multiplying it with a reweighting factor $p(\mathbf{a}|\boldsymbol{\rho})/p(\mathbf{a}|\boldsymbol{\rho}_0)$. This is the main result of this work. The reweighting factor is a ratio of two Gaussian distributions, which can be computed in closed form. Moreover, since all of the terms are Gaussian in $\mathbf{a}$, we can marginalize over $\mathbf{a}$ to obtain the posterior distribution of the hyperparameters $\boldsymbol{\rho}$ analytically. For a full array, of pulsars, the full posterior distribution becomes

$$
p(\mathbf{a},\boldsymbol{\rho}|\delta\mathbf{t}) \approx \prod_{k=1}^{N_p}\left[\frac{\mathcal{N}(\mathbf{a}_k|\hat{\mathbf{a}}_{0,k},\Sigma_{0,k})}{p(\mathbf{a}_k|\boldsymbol{\rho}_{0,k})}\right]p(\mathbf{a}|\boldsymbol{\rho})p(\boldsymbol{\rho}), \quad (21)
$$

where we adopt the convention that the subscript $k$ refers to the $k$th pulsar. When no subscript is present, the quantity $\mathbf{a}$ or $\boldsymbol{\rho}$ refers to the concatenation of the parameter for all the pulsars.

An intuitive interpretation of Eq. (21) can be given in terms of a likelihood function in Fourier space. The quantity $\mathcal{N}(\hat{\mathbf{a}}_{0,k}|\mathbf{a}_k,\Sigma_{0,k})$, where we swapped the position of $\hat{\mathbf{a}}_{0,k}$ and $\mathbf{a}_k$, can be thought of as a likelihood function

where $\hat{\mathbf{a}}_{0,k}$ now takes the role of the data. Indeed, we see in Eq. (17) that $\hat{\mathbf{a}}_{0,k}$ is the quantity that depends on the data $\delta\mathbf{t}$ (regularized by the ridge regression prior). Marginalizing Eq. (21) over $\mathbf{a}$ gives us the posterior distribution of the hyperparameters $\boldsymbol{\rho}$

$$p(\mathbf{a},\boldsymbol{\rho}|\delta\mathbf{t}) \approx \prod_{k=1}^{N_p}\left[\frac{\mathcal{N}(\hat{\mathbf{a}}_{0,k}|\mathbf{a}_k,\Sigma_{0,k})}{p(\mathbf{a}_k|\boldsymbol{\rho}_{0,k})}\right]p(\mathbf{a}|\boldsymbol{\rho})p(\boldsymbol{\rho})$$

$$p(\boldsymbol{\rho}|\delta\mathbf{t}) \approx \frac{\mathcal{N}(\hat{\mathbf{a}}_0|0,\Sigma_0)}{\mathcal{N}(\hat{\mathbf{a}}|0,\Sigma)}\sqrt{\frac{\det(2\pi\phi_0)}{\det(2\pi\phi)}}p(\boldsymbol{\rho}), \tag{22}$$

where $\hat{\mathbf{a}}$ and $\Sigma$ are functions of $\boldsymbol{\rho}$

$$\Sigma^{-1} = \Sigma_0^{-1} + \phi^{-1} - \phi_0^{-1}$$
$$\hat{\mathbf{a}} = \Sigma\Sigma_0^{-1}\hat{\mathbf{a}}_0, \tag{23}$$

with $\Sigma_0$ the block-diagonal concatenation of the collection of $\Sigma_{0,k}$ of all pulsars, and $\mathbf{a}_0$ the concatenation of all the $\mathbf{a}_{0,k}$. Equation (22) defines the full-PTA posterior distribution as a function of the results of the output of our *Step 1*. In most contemporary PTA analyses, the white noise parameters $\boldsymbol{\theta}$ are typically held fixed for computational efficiency. Although it is formally not correct, because the $\boldsymbol{\theta}$ are assumed to not be covariant with $\boldsymbol{\rho}$ and $\mathbf{a}$, holding $\boldsymbol{\theta}$ fixed to their maximum likelihood estimators of a single-pulsar analysis is deemed sufficiently accurate for a full PTA analysis. When holding $\boldsymbol{\theta}$ fixed, Eq. (20) becomes an identity, and consequently Eq. (22) becomes identical to the distribution that is implemented in packages like ENTERPRISE. With our method, we can now vary $\boldsymbol{\theta}$ during *Step 1*.

## III. APPLICATION AND EXAMPLES

In this section, we discuss the implementation of the method described in Sec. II C when carrying out Bayesian inference runs over a PTA dataset. We use the 25 pulsars of the EPTA DR2new dataset [42] and compare our results with the results obtained with the "standard" likelihood [Eq. (11)] coded in ENTERPRISE [36]. We first discuss the example of a SPNA for the pulsar J1738+0333, and then show the results of a GWB search on the full pulsars array.

### A. Pulsar J1738 + 0333: An example of SPNA

The timing solution of the pulsar J1738+0333 is based on data collected with the Arecibo and the EPTA telescopes. For the analysis described in this work, we use the EPTA DR2new data [49]. The noise analysis carried out by the EPTA collaboration is described in [50]. Evaluating the Bayes factor between different noise models, they show that in the DR2new dataset a noise model with a chromatic noise-only is slightly favored with respect to a model that contains both chromatic and achromatic noise contributions.

TABLE II. Prior distributions as defined for the presented analysis.

| | Free parameter | Prior type | Interval |
|---|---|---|---|
| *Step 1* | EFAC ($\mathcal{E}$) | Uniform | [0.5, 5] |
| | EQUAD ($\mathcal{Q}$) | Log-uniform | [−10, −5] |
| | $\log_{10}A_{\mathrm{DMdip}}$ | Log-uniform | [−10, −2] |
| | $\log_{10}\tau_{\mathrm{DMdip}}$ | Log-uniform | [0, 2.5] |
| | $t0_{\mathrm{DMdip}}$ | Uniform | [$t_{\mathrm{min}}$, $t_{\mathrm{max}}$] |
| *Step 2* | $\log_{10}A_{\mathrm{RN/DM/chrom}}$ | Log-uniform | [−18, −12] |
| | $\gamma_{\mathrm{RN/DM/chrom}}$ | Uniform | [0, 7] |
| | $\log_{10}k_{\mathrm{RN/DM/chrom}}$ | Log-uniform | [−9, −4] |
| | $\log_{10}A_{\mathrm{GWB}}$ | Log-uniform | [−15.5, −13.5] |
| | $\gamma_{\mathrm{GWB}}$ | Uniform | [0, 7] |

We present here the results for a SPNA carried out on J1738+0333 (EPTA DR2new data) with our regularized likelihood. We defined a noise model that includes EFAC and EQUAD (which model the TOAs measurements uncertainty, white noise) specific for each observing system, and RN and DM variations, both modeled as stationary Gaussian processes with a *flat-tail power-law spectrum*

$$S_{\mathrm{y}}(f;A_{\mathrm{y}},\gamma_{\mathrm{y}},k_{\mathrm{y}}) = \max\left(\frac{A_{\mathrm{y}}^2}{12\pi^2}\left(\frac{f}{\mathrm{yr}^{-1}}\right)^{-\gamma_{\mathrm{y}}}\Delta f \mathrm{yr}^3, k_{\mathrm{y}}^2\right), \tag{24}$$

where $\Delta f$ is the inverse of the observation time and the subscript "y" refers either to RN or DM variations. We denote the timespan of one year as $yr$, which makes the units of the $S_{rmy}$ as used in ENTERPRISE ($sec^2$). The reasoning behind using a flat-tail power-law instead of a simple power-law model is discussed in Appendix A. We model RN and DM over, respectively, 30 and 100 frequency bins.

We follow the methodology described in Sec. II C to obtain posterior distributions for the RN and DM variations hyperparameters. We used the Markov chain Monte Carlo (MCMC) sampler PTMCMCSampler [51]. First, we sampled over the white noise parameters using the time-domain likelihood [Eq. (11)], while setting the RN and DM hyperparameters fixed[3] to $\log_{10}A = -12$, $\gamma = 5$ and $\log_{10}k = -5$ (*Step 1*). We used the results to estimate $\Sigma_0$ and $\hat{\mathbf{a}}_0$ as described in Eq. (19), and then run inference over the RN and DM hyperparameters using Eq. (21) (*Step 2*). A summary of the prior distributions considered for these analyzes can be found in Table II.

[3]Note that the only constraints on the choice of $\boldsymbol{\rho}_0$ are given by the numerical stability of the algorithm. Such a high $\log_{10}k$ ensures that $\phi_0^{-1}$ is smaller than $\phi^{-1}$ [Eq. (20)] for every $\boldsymbol{\rho}$, and the final covariance matrix is positive definite. More details are discussed in Appendix. A.

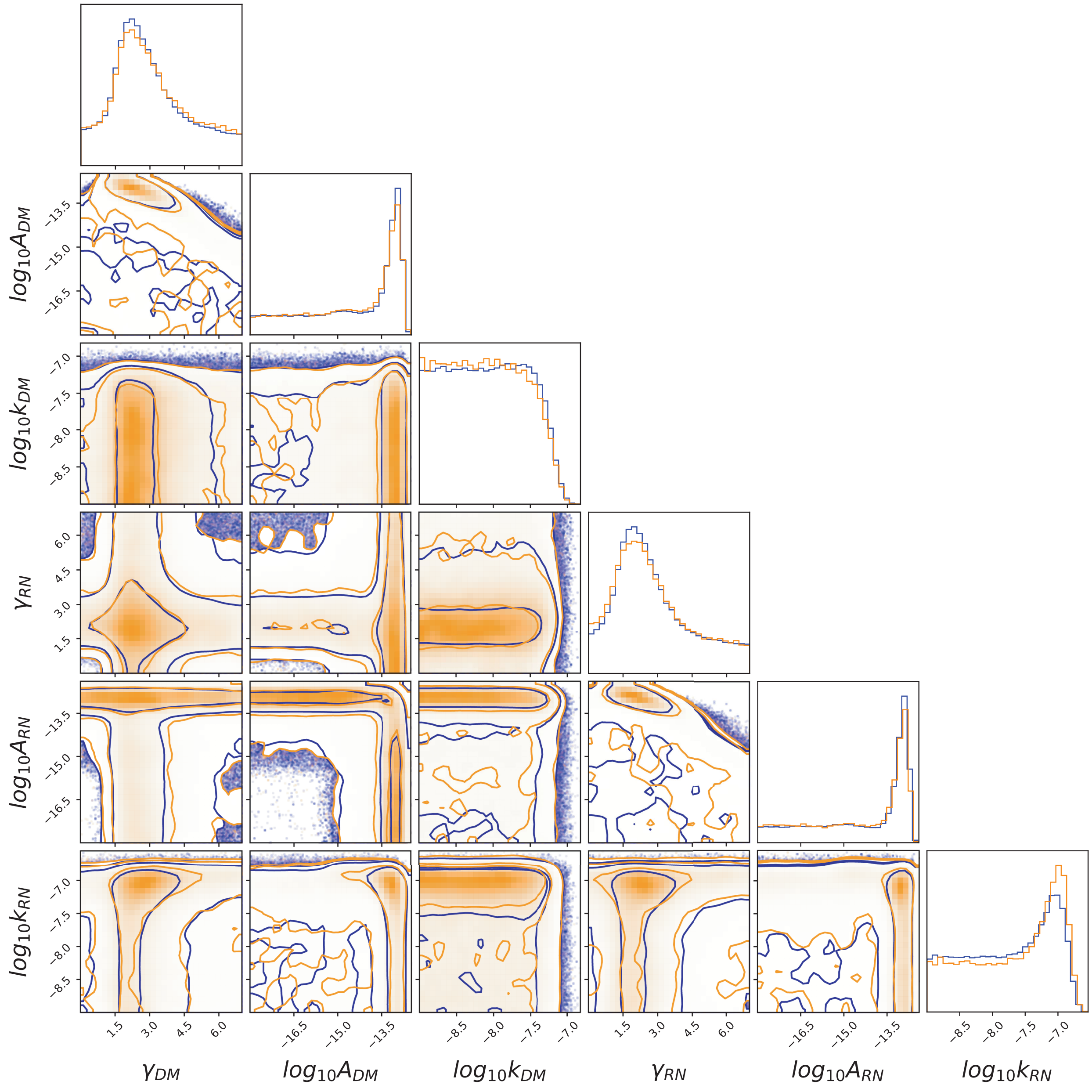

FIG. 1. Posteriors for J1738 + 0333 noise analysis. The orange posteriors are obtained for an inference run over the Gaussian noise processes with the regularized likelihood in Fourier domain; the blue posteriors come from a full SPNA over all the noise processes using the time-domain likelihood.

We show in Fig. 1 the posteriors obtained for the RN and DM variations hyperparameters with the regularized likelihood (orange curves). Those are compared with the posteriors obtained for a full SPNA using the traditional likelihood [Eq. (11), blue curve]. The RN and DM posteriors obtained with our method (*Step 1* + *Step 2*) are equivalent to what one would obtain sampling over all the noise parameters (including the white noise parameters we marginalize over in *Step 1*) with the time-domain likelihood. The slight widening of the orange posteriors, compared to the blue posteriors, is due to the fact that we are marginalizing over the white noise parameters. Even when white noise parameters and GWB hyperparameters are not covariant, marginalizing over the white noise parameters is statistically more correct than fixing them to the maximum likelihood values obtained from the SPNA, and will result in slightly wider posteriors for the GWB hyperparameters too. A tutorial about reproducing Fig. 1 can be found at [52].

From Fig. 1, we can also notice that the obtained posteriors for the RN and DM hyperparameters are highly covariant. In Appendix C, we discuss a possible method we tried out to capture the covariance between RN and other signals using principal component analysis (PCA) principles. Being able to discriminate between RN and DM variations confidently would allow us to include the DM hyperparameters in *Step 1*. Unfortunately we did not get satisfactory results, so we include Appendix C only for reference (and because we like the approach).

In this section, we showed the results obtained for the pulsar J1738+0333. We also tested all the other 24 pulsars of the EPTA `DR2new` and always obtained excellent consistency between the RN and DM hyperparameters posteriors obtained with the time-domain likelihood and the regularized likelihood in Fourier domain. For the case of J1600−3053, the EPTA `DR2new` noise analysis [50] found evidence for an additional chromatic noise component. This is also supported by our method and can be sampled for in *Step 2*. In the case of the DM dip found in the J1713 + 0747 data, this deterministic signal can also be included in our model: a DM dip is not covariant with the GWB and it is not modeled as a Gaussian process in Fourier space; thus, it becomes one of the signals that get marginalized over after *Step 1*.

### B. GWB search

In this section, we present the results obtained carrying out a GWB search on EPTA `DR2new` dataset with our regularized likelihood in Fourier domain [Eq. (21)] and compare them with the results obtained with the standard time-domain likelihood [Eq. (11)].[4] As expected, the two methods are perfectly equivalent [see posteriors in Fig. 2: the blue posteriors are obtained from the standard Bayesian inference study in the time domain (Sec. II B), while the orange posteriors are obtained with the Fourier-domain method described in Sec. II C].

To obtain the posteriors in Fig. 2, we first carried out *Step 1* analysis individually on all the 25 pulsars of EPTA `DR2new`. For each of them, we obtained samples of the white noise parameters (EFAC and EQUAD specific for each observing system). At the same time, the RN and DM variations hyperparameters were fixed to $\log_{10} A = -12$, $\gamma = 5$, and $\log_{10} k = -5$. RN and DM noise processes are modeled over, respectively, 30 and 100 frequency bins for all pulsars. Additionally, according to the results of the customized noise analysis carried out in [50], an exponential DM dip (deterministic signal) was added for J1713 + 0747 and sampled over in *Step 1*. Furthermore, the noise model of J1600−3053 includes an additional chromatic noise signal modeled as a Gaussian process (with a flat-tail power-law spectrum) and whose hyperparameters are fixed in this first analysis to $\log_{10} A = -12$, $\gamma = 5$, and $\log_{10} k = -5$.

From the results of those single pulsar analyzes, we obtain an estimate of $\Sigma_0$ and $\hat{\mathbf{a}}_0$, as described in Eq. (19). We now have all the elements to use the posterior distribution function of Eq. (21) to conduct an inference run over all the pulsars' Gaussian noise processes and the GWB hyperparameters. We assumed the GWB to be stationary, Gaussian and with a power-law spectrum. The GWB amplitude and slope posteriors are shown in Fig. 2 (orange curve).

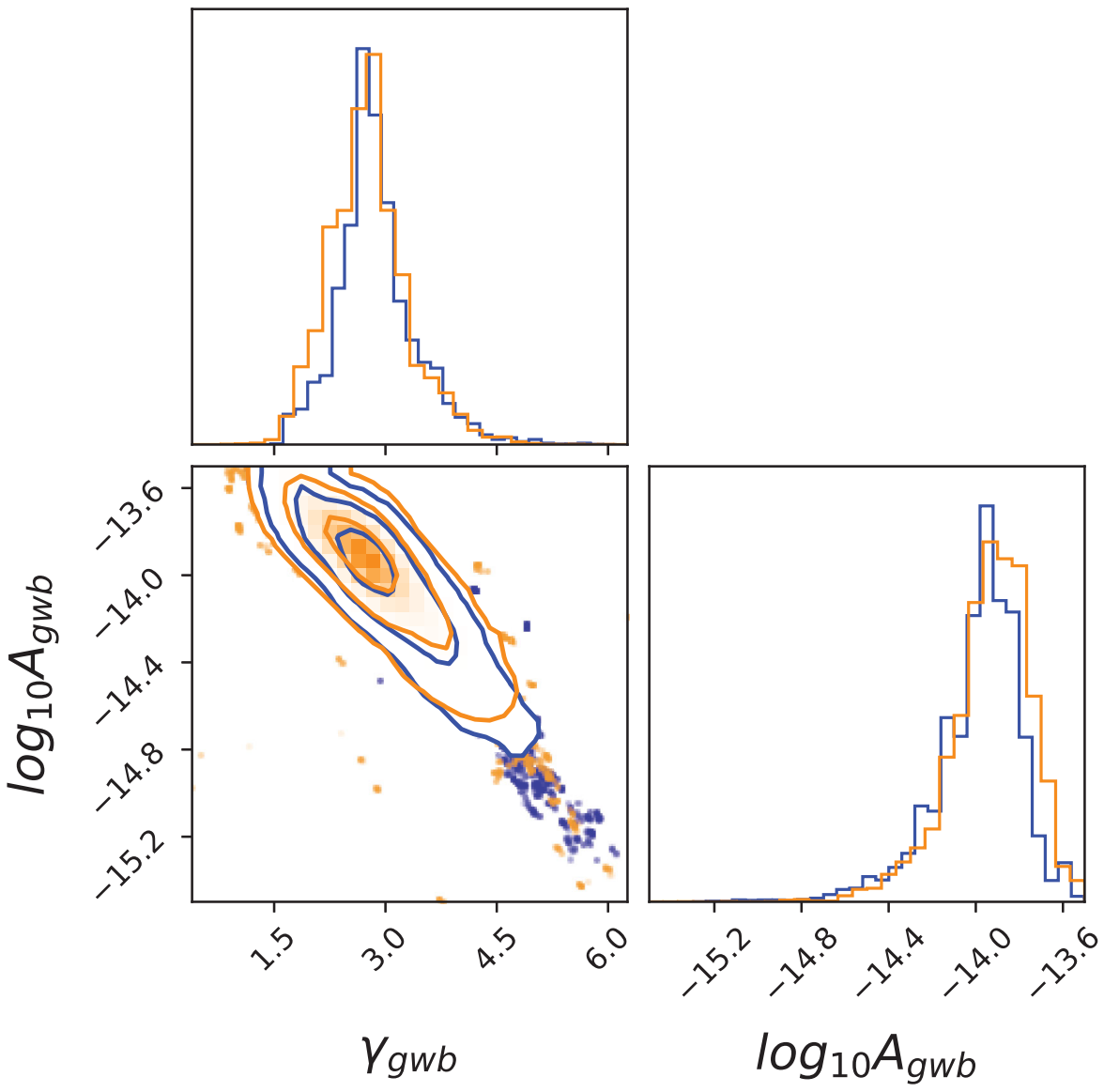

FIG. 2. Posteriors for the GWB hyperparameters obtained from the 25 pulsars of the EPTA `DR2new` dataset. The orange posteriors are obtained with the regularized likelihood in Fourier domain (the WN and DM dip parameters are marginalized over); the blue posteriors are obtained with the time-domain likelihood (the WN parameters are fixed).

From Fig. 2 it is clear that the results obtained with the time-domain likelihood and the regularized likelihood in Fourier domain are equivalent. However, the way white noise is included in the analysis is different. In the time-domain analysis, the white noise parameters are fixed to the maximum likelihood values obtained in the SPNA runs. In the Fourier-domain analysis, instead, we marginalized over the white noise parameters in *Step 1*. Furthermore, we are able to marginalize over other deterministic signals, like exponential DM dips.

The code used to produce the posteriors in Fig. 2 can be found at [52].

### C. Future applications: Gamma-ray PTA

In the first gamma-ray PTA data release [26], the results from two different strategies to fit for the timing model and RN parameters from gamma-ray pulsars' data were presented. The first method is the same method used for radio pulsars: a continuous observation of the pulsar is "folded" and averaged over the pulsar spin period; the resulting peak is cross-correlated with a template to obtain a TOA. While this method is very efficient for radio pulsars, for many of the observed gamma-ray millisecond pulsars, the limited

[4]The results of the GWB search on the EPTA `DR2new` dataset carried out by the EPTA collaboration can be found in [9,53].

*exposure* (collecting area per time) would make it necessary to fold many months of data in order to build a constrained TOA. An alternative method is the photon-to-photon approach [30]. In this case, each photon gets assigned a probability ($w_i$) that it belongs or not to the pulse template of that pulsar; those probabilities are used as weights in the pulsar likelihood to determine the timing model parameters. Contrarily to the likelihood function used for radio PTA [Eq. (11)], this likelihood is not a Gaussian distribution. Thus, it can be computed for any sample of values for the timing model and signal processes parameters, but it is not possible to analytically marginalize over any parameter because we do not have a conjugate prior for those parameters. [Note that, in standard radio PTA inference, we *almost always* marginalize over the timing model parameters and the Fourier coefficients used to describe the included Gaussian processes (an example of study where this marginalization does not occur is [47]).]

Given a set of photon observations and the corresponding weights, we can write the Poissonian likelihood for the data $D$ as a function of the timing model parameters $\boldsymbol{\beta}$, the Fourier coefficients $\mathbf{a}$ and the template pulse profile parameters $\boldsymbol{\tau}$ as

$$p(D|\boldsymbol{\beta},\mathbf{a},\boldsymbol{\tau}) = \prod_i w_i p(\Phi_i|\boldsymbol{\tau}) + (1 - w_i), \qquad (25)$$

where $p(\Phi_j|\boldsymbol{\tau})$ is the template pulse profile (usually written a sum of symmetrically wrapped Gaussian peaks) evaluated at the phase $\Phi_i = \Phi(t_i;\boldsymbol{\beta},\mathbf{a})$ for the $i$th photon. The phase model $\Phi(t_i;\boldsymbol{\beta},\mathbf{a})$ is the sum of the predicted phase from the timing model solution at the time $t_i$ and the description in the Fourier basis of additional noise processes. In Gamma-ray PTA pulsar noise inference, the parameters searched for are both the timing model parameters and the Fourier coefficients of the waveform describing the Gaussian processes involved (remember that, in this case, there is no DM dispersion effect). More details about this type of analysis can be found at [26,54].

To date, the photon-to-photon approach was missing a method to look for correlated signals among different pulsars. In the first Fermi PTA data release [26], the photon-to-photon approach was used to infer single-pulsar upperlimits on the GWB amplitude. Since there is no cross-pulsar correlated term included in this method, the joint limit was obtained by multiplying the single pulsar posteriors and integrating the resulting distribution. One of the main motivations for this work was to derive a PTA likelihood function that could be applied to gamma-ray PTA data obtained with the photon-to-photon approach. The posterior samples of the Fourier coefficients (describing the Gaussian signal processes in each pulsars' data) can be used as input for our regularized likelihood [Eq. (21)] when looking at the whole array. We leave this analysis for future work.

## IV. CONCLUSIONS

We presented a regularized formulation of the PTA likelihood in Fourier domain [Eq. (21)] and showed results for both single pulsar noise analysis (Fig. 1) and GWB searches (Fig. 2) using the EPTA `DR2new` dataset [49]. We proved that our formulation is analytically equivalent to the time-domain likelihood [Eq. (11)] when the white noise (and deterministic signals) parameters are held fixed.

Our regularized likelihood in Fourier domain can be seen as the product of two terms [Eq. (21)]: (a) a Gaussian distribution in the Fourier coefficients $\mathbf{a}$, describing all the Gaussian processes included in the model, which depend on a given set of model hyperparameters $\boldsymbol{\rho}_0$; (b) a reweighting term consisting in the ratio between the prior probabilities evaluated for a general sample of the model hyperparameters $\boldsymbol{\rho}$ and for $\boldsymbol{\rho} = \boldsymbol{\rho}_0$. (See Sec. II C for more details.)

We identify two main advantages coming from using our regularized likelihood function [Eq. (21)]:

(i) It allows the splitting of the GWB inference on a PTA dataset in a two-step analysis. *Step 1* consists of conducting inference on the parameters of those signals not modeled in Fourier space (white noise and deterministic signals); each pulsar is analyzed individually. *Step 2* analyses the whole array simultaneously and produces inference on the hyperparameters of the GWB and signals covariant with it (RN, DM variations, etc.). All the parameters investigated in *Step 1* are marginalized over. Marginalizing over the noise parameters of the signals not covariant with the GWB brings also a reduction of the computational cost for an analysis on the full array of pulsars. Ideally, we would be able to include all signal processes in *Step 1*, except for the GWB and pulsars intrinsic RNs, reducing the computational cost to the minimum. In Appendix C we present a possible approach to include DM variations in *Step 1*; unfortunately, this method fails to fully describe the covariance between RN and DM hyperparameters because of the poor frequency coverage of the data.

(ii) It favors the immediate inclusion of gamma-ray data in the PTA dataset alongside the radio timing data without waiving the complexity and pulsar-specificity of the radio noise models. Analyzing gamma-ray data with the *photon-to-photon* approach [30], we evaluate the pulse phase at each individual photon-time and compare it with a template. We can then write a Poisson likelihood for this data that fits both the timing model parameters and the Fourier coefficients of the involved noise components (modeled as Gaussian processes, [54]). The resulting Fourier coefficients become the input of our PTA likelihood [Eq. (21)] when carrying out inference studies on the whole array. Thus, this regularized likelihood can be applied to both radio and gamma-ray data, independently of the software and models used to obtain the Fourier coefficients from the raw data. This opens the way for possible direct comparisons between the results from the two datasets.

We also made use of NumPy [55], Matplotlib [56], CORNER plot [57], SciPy [43], TEMPO2 [34,35], LIBSTEMPO [58], ENTERPRISE [36], PTMCMCSampler [51] and LA_FORGE [59].

## ACKNOWLEDGMENTS

We thank Colin Clark for insightful discussions about gamma-ray pulsar observations and data analysis and Michele Vallisneri for helpful comments on the first version of this paper. This work was supported by the Max Planck Gesellschaft (MPG) and the ATLAS cluster computing team at Albert Einstein Institute (AEI) Hannover.

## DATA AVAILABILITY

The data that support the findings of this article are openly available [42].

## APPENDIX A: FOURIER COEFFICIENTS SAMPLING AND NUMERICAL RESOLUTION PROBLEMS

The main idea that inspired our paper is that of the Fourier transform: our goal was to carry out an analysis similar to a Fourier transform, so that we could work with the Fourier-transformed data. Such an approach would make it easier to combine radio, gamma-ray, or other types of pulsar timing data. Moreover, this could then potentially be done on a per-pulsar basis, after which a full PTA analysis would be done on the data product of the per-pulsar analysis. The methods presented by [60] can be seen as such an effort. However, this is difficult to achieve in full generality, because the pulsar timing data is complex: the data is sampled irregularly, data products are not the same for radio/gamma-ray data, the signals of interest are very low-frequency, and we have to take into account the effects of the timing model. Let us briefly review the Fourier transform here in order to motivate our regularization technique in detail.

### 1. Regularization

In classical time-series analysis, the data can be converted from the time-domain to the Fourier domain by the discrete Fourier transform (DFT): an invertible linear transformation of the data which we can write as

$$\delta t = F\tilde{\delta t}, \tag{A1}$$

where $\tilde{\delta t}$ is the DFT of the data $\delta t$. In Eq. (A1), the transformation matrix $F$ has complex exponential basis functions as its columns, with frequencies being multiples of the fundamental harmonic of the time-series. Fast implementations of Eq. (A1) exist in the form of the fast Fourier transform [61].

The DFT is an invertible transformation, meaning that the columns of $F$ constitute a complete basis for our time series $\delta t$. This also means that both the signal and the noise are fully represented by $\tilde{\delta t}$. Often times in signal processing, including in PTA data analysis, the goal is to capture the underlying processes of interest, not the noise. Thinking of $\tilde{\delta t}$ from that perspective, the fact that the DFT is invertible means that we are *overfitting* the data. Overfitting is a consequence of having too many degrees of freedom in the model, resulting in capturing not only the processes of interest, but also the noise. Common solutions to overfitting include (1) reducing the degrees of freedom in the model, or (2) regularizing the model using constraints. In the machine learning literature, many regularization techniques have been developed over the past decades [62].

In pulsar timing, it is customary (for computational reasons) to use fewer frequencies, meaning that $F$ becomes a nonsquare matrix with fewer columns than rows. This means that the linear system of Eq. (A1) can be interpreted as an overdetermined system, and solutions like those obtained through least-squares optimization can be used. As hinted at above, reducing the number of frequencies is a regularization technique, because it removes degrees of freedom of the model. Indeed, if the higher frequencies are omitted from the model, the model should only capture trends of lower frequency.

In pulsar timing the data is sampled irregularly and quadratic spindown always needs to be taken into account when doing spectral analysis [63]. The result is that all Fourier transform elements are covariant. Moreover, there is degeneracy between the quadratic spindown parameters and the lowest frequency Fourier coefficients if we allow all those parameters to be free unconstrained in the model. This degeneracy can be broken by placing a constraint on the Fourier coefficients. We do this by placing a Gaussian prior on the Fourier coefficients. This regularizes the posterior distribution and breaks the degeneracy with the timing model. And, because we have an *analytical* description of the Gaussian prior, we can *undo* it at a later stage in the analysis without running into numerical issues stemming from finite number of samples (a usual problem with resampling approaches). This approach retains all information in the data.

### 2. Numerical resolution

The $\boldsymbol{\rho}_0$ define the regularization of the Fourier coefficients that we use in Sec. II C. Care needs to be taken in choosing the values of $\boldsymbol{\rho}_0$ so that numerical issues like under/overflow and roundoff errors are avoided. The distribution of Fourier coefficients is obtained by estimating directly its mean and variance, rather than computing them from $\mathbf{a}$ samples. The $\Sigma_0$ matrix is evaluated from the distribution of $\Sigma_{0i}$ matrices, all of which are obtained from a given set of samples of all the hyperparameters and, thus, are not singular. In our analysis, we used one thousand of $\Sigma_{0i}$ samples to compute $\Sigma_0$ using the update formula in Eq. (19). This resulted in an accuracy on the estimator the individual elements of the covariance matrix of the order of 0.04%.

An alternative method to estimate the covariance matrix $\Sigma_0$ would be to compute it directly from a set of samples of the Fourier coefficients $\mathbf{a}$ describing the Gaussian processes involved in the model [we use that approach in Appendix C, see Eqs. (C10) and (C11)]. This may lead to numerical resolution problems when computing the inverse of $\Sigma_0'$.

The $\Sigma_0'$ matrix is characterized by both very large and very small eigenvalues. Even if most of the covariance information lies in the larger eigenvalues, the small eigenvalues are necessary to guarantee the matrix is positive definite. $\Sigma_0'$ has a very high condition number and is a so-called *ill-conditioned* matrix. To guarantee numerical stability, the number of samples should be at least of the same order of magnitude as the product between the matrix's condition number and the number of features $m'$ [$\Sigma_0'$ is a $(m' \times m')$ matrix] [64].

The condition number of $\Sigma_0'$ can be reduced for optimal choices of the $\boldsymbol{\rho}_0$ values to which the Gaussian process hyperparameters are fixed. From looking at the reweighting term in

$$p(\mathbf{a},\boldsymbol{\rho}|\delta\mathbf{t}) \approx \mathcal{N}(\mathbf{a}|\hat{\mathbf{a}}_0,\Sigma_0)\frac{p(\mathbf{a}|\boldsymbol{\rho})}{p(\mathbf{a}|\boldsymbol{\rho}_0)}p(\boldsymbol{\rho}), \tag{A2}$$

it is clear that the distribution at the denominator has to be wider than the condition number at the numerator, such that the ratio is still a Gaussian distribution. Thus, the $\boldsymbol{\rho}_0$ values should be chosen accordingly. Assuming a flat-tail power-law spectra for both RN and DM variations favors the stability of the algorithm. In fact, by comparing the periodograms (spectral density plots as a function of frequency) of those processes for different values of the $\boldsymbol{\rho}_0$ parameters (for a flat-tail power law spectra: $\boldsymbol{\rho}_0 = [\log_{10}A_0, \gamma_0, \log_{10}k_0]$), it becomes clear that the $\log_{10}k_0$ is the real discriminant in this equation. In practice, for realistic PTA datasets, a value of $\log_{10}k \in (-6,-3)$ should be enough to guarantee that the denominator distribution is wider than the distribution at the numerator, making the posterior is normalizable. Although, the exact range is pulsar specific and depends on the observation cadence. We have found no problems with $\log_{10}k = -5$. Other potential solutions are standard matrix regularization or Cholesky rank-1 update algorithms.

## APPENDIX B: COVARIANCE MATRIX UPDATE FORMULA

In this appendix, we derive the formula used to compute the variance $\Sigma_0$ from the distribution of $\Sigma_{0i}$ elements [Eq. (19)]. We start from a easier case, where we want to write the covariance matrix of a set of $n+p$ samples ($C_{n+p}$) as a function of the covariance matrices $C_n$ and $C_p$ (obtained considering, respectively, only the first $n$ and the last $p$ samples), and then generalize the result to obtain Eq. (19).

Given $n$ independent observations (samples) of the set of parameters $\mathbf{x}$, where $\mathbf{x} = (x_1, x_2, \ldots x_m)$, the covariance matrix $C_n \in \mathcal{R}^{n\times m}$ is defined as

$$\begin{aligned} C_n &\equiv \frac{1}{n-1}\sum_{i=1}^{n}(\mathbf{x}_i-\hat{\mathbf{x}}_n)(\mathbf{x}_i-\hat{\mathbf{x}}_n)^T \\ &= \frac{1}{n-1}\left[\sum_{i=1}^{n}\mathbf{x}_i\mathbf{x}_i^T - \sum_{i=1}^{n}\mathbf{x}_i\hat{\mathbf{x}}_n^T \right.\\ &\qquad \left. - \hat{\mathbf{x}}_n\left(\sum_{i=1}^{n}\mathbf{x}_i\right)^T + n\hat{\mathbf{x}}_n\hat{\mathbf{x}}_n^T\right] \\ &= \frac{1}{n-1}\left[\sum_{i=1}^{n}\mathbf{x}_i\mathbf{x}_i^T - n\hat{\mathbf{x}}_n\hat{\mathbf{x}}_n^T\right], \end{aligned} \tag{B1}$$

where $\hat{\mathbf{x}}_n = \sum_i \mathbf{x}_i/n$ is the mean over the $n$ samples of $\mathbf{x}$. Note that, by definition, when $n=1$ the fraction in Eq. (B1) becomes simply $1/n = 1$.

Consider now $n+p$ samples of $\mathbf{x}$. The mean $\hat{\mathbf{x}}_{n+p}$ can be rewritten as

$$\hat{\mathbf{x}}_{n+p} = \frac{1}{n+p}\left(\sum_{i=1}^{n}\mathbf{x}_i + \sum_{j=1}^{p}\mathbf{x}_j\right) = \frac{1}{n+p}(n\hat{\mathbf{x}}_n + p\hat{\mathbf{x}}_p). \tag{B2}$$

From Eq. (B1), it is immediate to see that the covariance matrix $C_{n+p}$ is defined as

$$\begin{aligned} C_{n+p} = \frac{1}{n+p-1}&\left[\sum_{i=1}^{n+p}\mathbf{x}_i\mathbf{x}_i^T - \sum_{i=1}^{n+p}\mathbf{x}_i\hat{\mathbf{x}}_{n+p}^T\right. \\ &\left. - \hat{\mathbf{x}}_{n+p}\left(\sum_{i=1}^{n+p}\mathbf{x}_i\right)^T + (n+p)\hat{\mathbf{x}}_{n+p}\hat{\mathbf{x}}_{n+p}^T\right]. \end{aligned} \tag{B3}$$

Dividing the sums $\sum^{n+p} = \sum^{n} + \sum^{p}$ and using Eq. (B2), we can rewrite Eq. (B3) as a function of the mean and covariance of the two separate $n$ and $p$ sample sets

$$\begin{aligned} C_{n+p} = \frac{1}{n+p-1}&\Big[(n-1)C_n + (p-1)C_p \\ &+ n\hat{\mathbf{x}}_n\hat{\mathbf{x}}_n^T + p\hat{\mathbf{x}}_p\hat{\mathbf{x}}_p^T - (n+p)\hat{\mathbf{x}}_{n+p}\hat{\mathbf{x}}_{n+p}^T\Big] \end{aligned} \tag{B4}$$

independently of the values of $n$ and $p$.

This can be generalized to the case of $N$ sets of $n_k$ samples with means $\hat{\mathbf{x}}_k$ and covariances $C_k$ ($k = 1, \ldots N$). The total covariance matrix $C$ computed from all the $N$ sets of samples can then be expressed as a function of the covariance matrices and means of the individual sets

$$C = \frac{1}{(\sum_{k=1}^{N} n_k) - 1}\left[\sum_{k=1}^{N}((n_k - 1)C_k + n_k\hat{\mathbf{x}}_k\hat{\mathbf{x}}_k^T) - \left(\sum_{k=1}^{N} n_k\right)\hat{\mathbf{x}}\hat{\mathbf{x}}^T\right], \quad (B5)$$

where $\hat{\mathbf{x}}$ is the mean over all the $\sum n_k$ samples. This equation is exactly what is reported in Eq. (19) for the computation of the $\Sigma_0$ matrix from the set of $N_s$ matrices $\Sigma_{0i}$. Note that $n_k = 1$ for each $\Sigma_{0i}$ matrix (they are computed from a single sample of the noise hyperparameters).

## APPENDIX C: PCA APPROACH TO MARGINALIZE OVER DM VARIATIONS

In this paper, we presented a regularized formulation of the PTA likelihood in Fourier-domain [Eq. (21)] as an alternative to the commonly used time-domain likelihood [Eq. (11)]. The computational cost of evaluating the two likelihood functions is comparable: the dimensions of the involved matrices and the number of times those need to be inverted are almost identical. The number of free parameters is also similar. The advantage of using our regularized likelihood in Fourier domain is that deterministic signals (like, for example, DM dips) and other signals not covariant with the GWB can be sampled over in *Step 1* of the analysis, and then marginalized over while carrying out the inference run over the whole array. This reduces the number of free parameters involved. (See Sec. II C for more details.)

In this appendix, we investigate the possibility of including more signals in *Step 1* and further reduce the number of parameters when analyzing the full array (*Step 2*). The achromatic RN hyperparameters of the individual pulsars are strongly covariant with the GWB, defined as common spatially correlated red noise, thus have to be sampled over alongside with the common process hyperparameters. The DM variations effect, instead, could theoretically be disentangled from the RN processes because of its frequency dependence. In fact, the DM variations signal (chromatic red noise) is due to the interaction of the pulsar radio emission with the ionized interstellar medium, the Solar System interplanetary medium and the Earth's ionosphere. These interactions lead to frequency-dependent delays in the observed TOAs [Eq. (5)]. Here, we describe a possible method to marginalize over the DM variations hyperparameters without losing information about their covariance with RN and GWB hyperparameters. We note upfront that we did not get satisfactory results with this approach. We merely describe what we tried in this appendix.

The main idea is to carry out a principal component analysis (PCA) to reduce the dimensionality of the problem and preserve information about the covariance between different hyperparameters. In our case, we want to capture the covariance between the Fourier coefficients describing the RN signal and the DM hyperparameters in a SPNA. PCA is based on the idea that the data—here, the Fourier coefficients—can be represented as a linear combination of a smaller number of uncorrelated variables: the principal components. The principal components are the eigenvectors of the covariance matrix of the data, and they are ordered by the amount of variance they explain. The first principal component explains the most variance, the second explains the second most, and so on.

So, given a matrix $X$ $(n \times p)$ (the data, stacked), each row of $X$ (a single sample of data) can be mapped to a new row vector of length $l < p$. Thus, in matrix notation, we can write $X' = XW_l$, where $X'$ $(n \times l)$ is the result of the PCA and contains roughly the same information as $X$. The matrix $W_l$ $(p \times l)$ is the orthogonal linear transformation that maps each element of $X$ in a new coordinate system of $l$ principal components. Calling $\mathbf{w}_j$ the $l$-dimensional unit vectors that constitute the row elements of $W_l$, $\mathbf{w}_1$ is

$$\mathbf{w}_1 \equiv \mathrm{argmax}_{(\|\mathbf{w}\|=1)}\{||X\mathbf{w}||^2\} = \mathrm{argmax}_{(\|\mathbf{w}\|=1)}\{\mathbf{w}^T X^T X\mathbf{w}\}. \quad (C1)$$

The $k$th component $(k > 1)$ can be computed by subtracting the $k - 1$ principal components from $X$

$$X_k = X - \sum_{j=1}^{k-1} X\mathbf{w}_j\mathbf{w}_j^T. \quad (C2)$$

Thus, the $k$th weight vector is computed as

$$\mathbf{w}_k = \mathrm{argmax}_{(\|\mathbf{w}\|=1)}\{\mathbf{w}^T X_k^T X_k\mathbf{w}\}. \quad (C3)$$

The rows of the matrix $W_l$ are the first $l$ principal components $\mathbf{w}_k$ $(k = 1, \ldots l)$. By construction, $\mathbf{w}_k$ are also the first $l$ right singular vectors of $X$ obtained by its singular value decomposition for the first $l$ singular values [$X = UCW^T$, with $C$ $(n \times p)$ rectangular diagonal matrix of the singular values, $U$ $(n \times n)$ and $W$ $(p \times p)$ contain, respectively, the left and right singular vectors]. Since

$$X^T X = WC^T U^T UCW^T = W\tilde{C}^2 W^T, \quad (C4)$$

where $\tilde{C}^2 \equiv C^T C$, the eigenvectors of $X^T X$ (covariance matrix between observed correlated variables) are the right singular vectors of the matrix $X$. See [65] for more details on PCA approaches.

In the following subsection, we show how to apply this method to our case of interest: a SPNA where we want to reduce the dimensionality of the problem without waving information on the covariance between different noise hyperparameters.

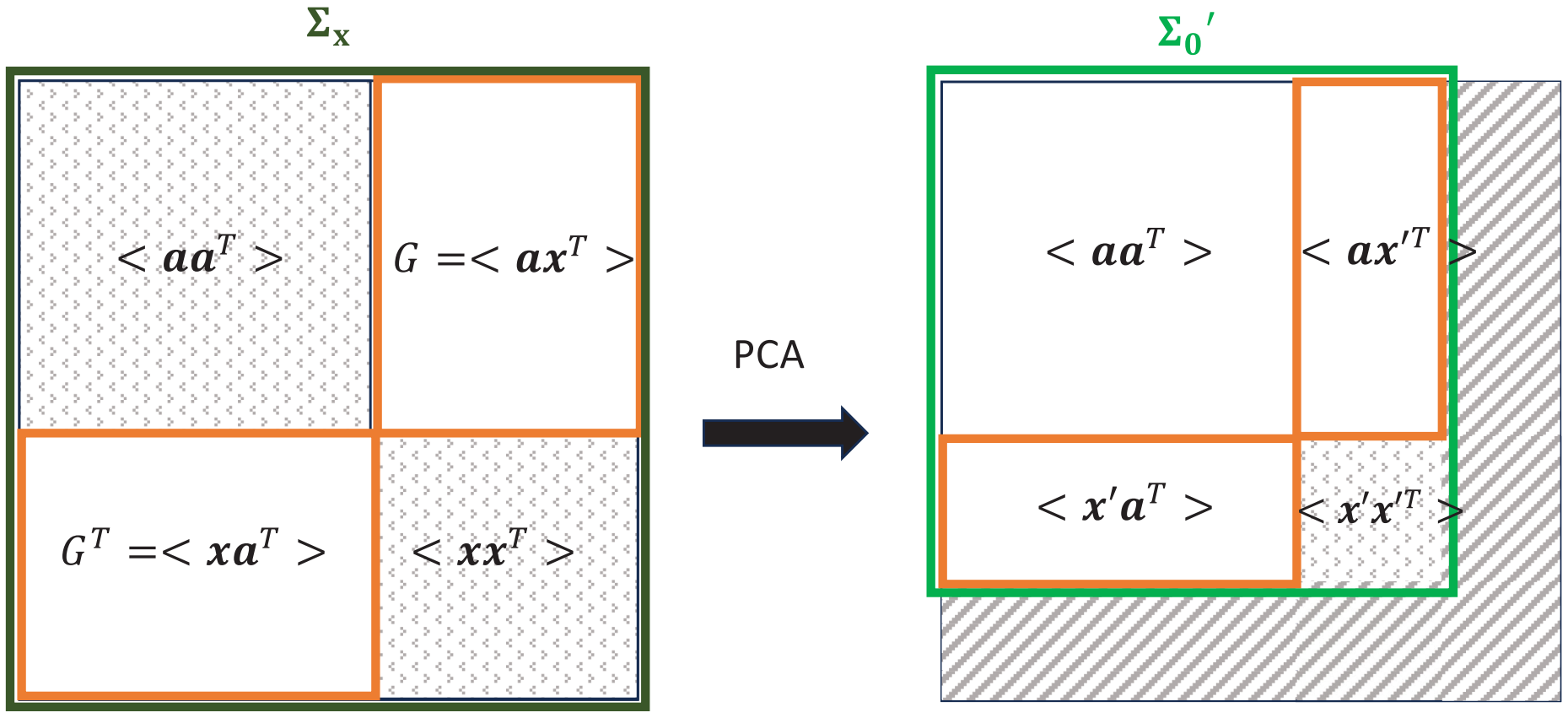

FIG. 3. Visualization of the PCA approach. The squares are schematic representations of the covariance matrices of $\Sigma_x$ [original set of variables, Eq. (C10)] and $\Sigma_0'$ [after PCA, Eq. (C11)]. $\mathbf{a}_{\mathrm{RN}}$ ($\mathbf{a}_{RN} = \mathbf{a}$ in this visualization) are the Fourier coefficients that describe the achromatic noise processes included in the model. $\mathbf{x}$ is the set of parameters investigated in *Step 1* [Eq. (C6)]. $\mathbf{x}'$ are optimized linear combinations of the parameters $\mathbf{x}$ [Eq. (C8)]. The parts of the matrices with the orange border contain the same covariance information. Note that $\langle \mathbf{a}\mathbf{a}^T \rangle$ can be computed analytically: $\langle \mathbf{a}\mathbf{a}^T \rangle = (F^T \tilde{N}^{-1} F + \phi^{-1})^{-1}$ [See Eq. (17)].

### 1. Implementation: Sampling over the Fourier coefficients

We describe here in detail the implementation of a PCA to marginalize over the DM variation hyperparameters when carrying out an inference run on PTA data. Notebook tutorials and the codes used to produce the figures in this appendix are available at [52]. A complete summary of the notation is in Table I. A schematic visualization of the method described in this section can be found in Fig. 3.

The main idea is to investigate the DM variations hyperparameters for each pulsar individually (include them in *Step 1*), and marginalize over them when analyzing the whole pulsars array without losing information on the covariance between the DM hyperparameters and the RN hyperparameters. In this scenario, Eq. (20) becomes

$$p(\mathbf{a}_{\mathrm{RN}}, \boldsymbol{\rho}_{\mathrm{RN}}, \boldsymbol{\rho}_{\mathrm{DM}} | \boldsymbol{\theta}, \delta\mathbf{t}) = p(\mathbf{a}_{\mathrm{RN}} | \boldsymbol{\theta}, \delta\mathbf{t}, \boldsymbol{\rho}_{\mathrm{DM}}, \boldsymbol{\rho}_{\mathrm{RN}_0}) \times \frac{p(\mathbf{a}_{\mathrm{RN}} | \boldsymbol{\rho}_{\mathrm{RN}}) p(\boldsymbol{\rho}_{\mathrm{RN}})}{p(\mathbf{a}_{\mathrm{RN}} | \boldsymbol{\rho}_{\mathrm{RN}_0})}, \tag{C5}$$

where the DM hyperparameters are no longer included in the reweighting term. Introducing

$$\mathbf{x} \equiv \begin{bmatrix} \boldsymbol{\theta} \\ \boldsymbol{\rho}_{\mathrm{DM}} \end{bmatrix}, \tag{C6}$$

a PCA is applied to rewrite $p(\mathbf{a}_{\mathrm{RN}} | \mathbf{x}, \boldsymbol{\rho}_{\mathrm{RN}_0})$ as $p(\mathbf{a}_{\mathrm{RN}} | \mathbf{x}', \boldsymbol{\rho}_{\mathrm{RN}_0})$, where $\mathbf{x}'$ is an optimized linear combination of $\boldsymbol{\theta}$ and $\boldsymbol{\rho}_{\mathrm{DM}}$ ($\mathbf{x}$ is an m-dimensional vector, $\mathbf{x}'$ is an m'-dimensional vector and $m' < m$) that preserves the covariance between those parameters and $\mathbf{a}_{\mathrm{RN}}$.

A normal distribution like $p(\mathbf{a} | \mathbf{x}, \boldsymbol{\rho}_{\mathrm{RN}_0})$ is fully described by its mean $\hat{\mathbf{a}}_0$ and variance $\Sigma_0$. The method described in Sec. II C directly estimates $\hat{\mathbf{a}}_0$ and $\Sigma_0$ from samples of the $\boldsymbol{\theta}$ parameters [Eqs. (17) and (19)]. The distribution $p(\mathbf{a}_{\mathrm{RN}} | \mathbf{x}', \boldsymbol{\rho}_{\mathrm{RN}_0}) = \mathcal{N}(\mathbf{a}_{\mathrm{RN}} | \hat{\mathbf{a}}_0', \Sigma_0')$ has mean $\hat{\mathbf{a}}_0'$ and variance $\Sigma_0'$. These two quantities not only describe the distribution of the Fourier coefficients $\mathbf{a}_{\mathrm{RN}}$ when the RN hyperparameters are set to $\boldsymbol{\rho}_{\mathrm{RN}_0}$, but also store information about their covariance with the DM hyperparameters $\boldsymbol{\rho}_{\mathrm{DM}}$. We describe here how to obtain the best estimates for these quantities.

*Step 1* is analogous to *step 1* described in Sec. II C, with the exception that DM hyperparameters are not fixed to some values $\boldsymbol{\rho}_{\mathrm{DM}_0}$, but sampled alongside the $\boldsymbol{\theta}$ parameters. The next step would be to evaluate the $\mathbf{x}'$ parameters in order to reduce the dimensionality of the problem without losing any information about the covariance between the hyperparameters of the involved signals.

We choose to evaluate the $\mathbf{x}'$ parameters with a PCA. Thus, $\mathbf{x}'$ are defined as a linear combination of the $\boldsymbol{\theta}$ and $\boldsymbol{\rho}_{\mathrm{DM}}$ parameters, which optimally describes the covariance between them and the Fourier coefficients describing RN $\mathbf{a}_{\mathrm{RN}}$. This information is encoded in the covariance matrix $G$:

$$G \equiv \langle \mathbf{a}_{\mathrm{RN}} \mathbf{x}^T \rangle. \tag{C7}$$

In order to evaluate $G$, we need samples of the Fourier coefficients $\mathbf{a}_{\mathrm{RN}}$. These can be drawn from the conditional distribution $p(\mathbf{a}_{\mathrm{RN}} | \mathbf{x}, \delta\mathbf{t})$ where the signal parameters $\mathbf{x}$ come from the analysis in *Step 1*. This can be done through e.g., the package LA_FORGE [59] or ENTERPRISE.[5]

[5]In Sec. II C we were able to compute $\hat{\mathbf{a}}_0$ and $\Sigma_0$ directly from the samples of the $\boldsymbol{\theta}$ parameters [Eqs. (17) and (19)], without drawing samples of the Fourier coefficients $\mathbf{a}$. In this case, we are interested in the covariance between $\mathbf{a}$ and $\mathbf{x}$, which cannot be estimated analytically from the $\boldsymbol{\theta}$ parameters samples only.

Given the covariance matrix $G$, we compute the eigenvalues and eigenvectors of the squared matrix $G^T G$. Since $G^T G$ has dimension $(m \times m)$ (where $m$ is the length of $\mathbf{x}$), we obtain as many eigenvalues as the number of $\boldsymbol{\theta}$ and $\boldsymbol{\rho}_{\mathrm{DM}}$ parameters. The larger the eigenvalue, the higher the covariance information encoded in the correspondent eigenvector. In particular, the "percentage of covariance information" contained in a subset of these eigenvectors is estimated as the percentage ratio between the sum of the correspondent subset of eigenvalues and the sum of all eigenvalues. Often, most of the covariance information ($> 99\%$) is encoded in very few eigenvectors. We set a threshold at 99% and select the lowest number of eigenvalues (and eigenvectors) for which 99% is represented. Defining $\mathbf{w}$ the matrix whose columns correspond to the selected eigenvectors, $\mathbf{x}'$ is defined as

$$\mathbf{x}' \equiv \mathbf{w}^T \mathbf{x}. \tag{C8}$$

Note that, by construction, the dimension of $\mathbf{x}'$ is smaller than the dimension of $\mathbf{x}$.

We can now define the transformation matrix $T_w$ as

$$T_w \equiv \begin{bmatrix} I_{n_{a\mathrm{RN}}} & 0 \\ 0 & \mathbf{w}^T \end{bmatrix}, \tag{C9}$$

where $I_{n_{a\mathrm{RN}}}$ is an identity matrix of dimension equal to the number of Fourier components used to describe RN. Given

$$\Sigma_x \equiv \begin{bmatrix} \langle \mathbf{a}_{\mathrm{RN}} \mathbf{a}_{\mathrm{RN}}^T \rangle & \langle \mathbf{a}_{\mathrm{RN}} \mathbf{x}^T \rangle \\ \langle \mathbf{x} \mathbf{a}_{\mathrm{RN}}^T \rangle & \langle \mathbf{x}\mathbf{x}^T \rangle \end{bmatrix} = \begin{bmatrix} \langle \mathbf{a}_{\mathrm{RN}} \mathbf{a}_{\mathrm{RN}}^T \rangle & G \\ G^T & \langle \mathbf{x}\mathbf{x}^T \rangle \end{bmatrix}, \tag{C10}$$

we can use the transformation defined in Eq. (C9) to compute the mean $\hat{\mathbf{a}}_0'$ and variance $\Sigma_0'$

$$\hat{\mathbf{a}}_0' \equiv \begin{bmatrix} \hat{\mathbf{a}}_0 \\ \mathbf{x}' \end{bmatrix}$$

$$\Sigma_0' \equiv T_w \Sigma_x T_w^T. \tag{C11}$$

From Eq. (C11), the distribution $p(\mathbf{a}_{\mathrm{RN}} | \mathbf{x}', \boldsymbol{\rho}_{\mathrm{RN}_0})$ is fully determined. Possible numerical resolution issues related to reconstructing the normal distribution from Fourier coefficients samples are discussed in Appendix A. Thus, we have all the elements to use Eq. (C5) and move to *Step 2* to sample over the RN hyperparameters.

Note that, for completeness, here we also included the white noise parameters in the PCA. This is not necessary, since white noise parameters are typically not covariant with the GWB. Thus, in the method described in Sec. II C it is not needed to add $\mathbf{x}'$ terms when deriving the normal distribution $p(\mathbf{a} | \delta\mathbf{t}, \boldsymbol{\rho}_0)$.

### 2. Results

We present the results obtained with the method presented in Appendix C 1 in the case of SPNA on the EPTA `DR2new` data [49]. As we mentioned before, we did not get satisfactory results with this approach for some pulsars. As an example, we show the results obtained with the PCA approach for the RN hyperparameters of one pulsar: J1738+0333.

We first carried out an inference run over the white noise and DM variation hyperparameters (*Step 1*) considering EFACs and EQUADs specific for each observing backend, and DM variations as a Gaussian process with a flat-tail power-law spectrum. The RN was also included in the model as a Gaussian process with a flat-tail power-law spectrum, but with the corresponding hyperparameters fixed to $\log_{10} A_{\mathrm{RN}} = -12$, $\gamma_{\mathrm{RN}} = 5$, and $\log_{10} k_{\mathrm{RN}} = -5$. Then, we carried out an inference run over the RN hyperparameters (*Step 2*) using the regularized likelihood [Eq. (21)], with $\hat{\mathbf{a}}_0'$ and $\Sigma_0'$ are derived as discussed in Appendix C 1 [Eq. (C11)]. See Fig. 4. The resulting posteriors (green curves) are compared with the posteriors obtained with the time-domain likelihood [Eq. (11)] for a full SPNA (blue curves).

In Fig. 4, the obtained blue (full SPNA with the time-domain likelihood) and green (Fourier-domain analysis with PCA approach to include the covariance between DM variations and RN hyperparameters) posteriors are still compatible, but the green posteriors are narrower than the blue posteriors, showing a loss of information about the RN hyperparameters. We expect the two sets of posteriors to be nearly identical for a method that is reliable. The discrepancy observed in Fig. 4 indicates that our PCA-based approach does not fully capture the covariance between RN and DM hyperparameters. Our best guess is that this is due to the little frequency coverage of pulsar J1738+0333 data, which does not provide enough information to successfully disentangle the two processes. This causes highly non-Gaussian posteriors with high covariance.

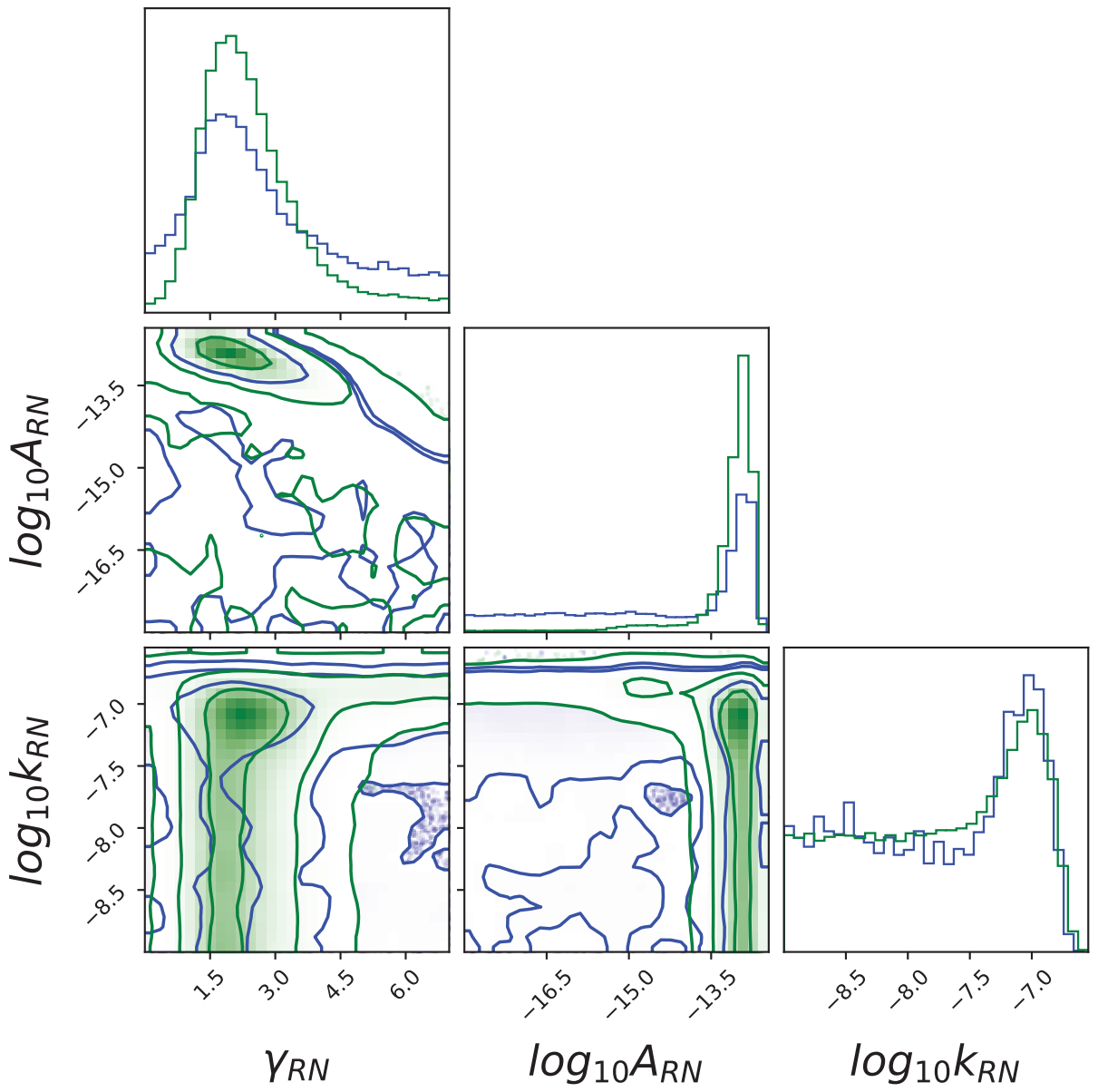

FIG. 4. Posteriors for J1738 + 0333 RN hyperparameters obtained with the PCA approach (green curves) described in Appendix C 1. The blue posteriors are instead the result of a full SPNA with the time-domain likelihood.

The codes used to obtain the results in Fig. 4 are available at [52].

[1] M. Rajagopal and R. W. Romani, Astrophys. J. **446**, 543 (1995).

[2] A. H. Jaffe and D. C. Backer, Astrophys. J. **583**, 616 (2003).

[3] J. S. B. Wyithe and A. Loeb, Astrophys. J. **590**, 691 (2003).

[4] A. Sesana, A. Vecchio, and C. N. Colacino, Mon. Not. R. Astron. Soc. **390**, 192 (2008).

[5] P. A. Rosado, A. Sesana, and J. Gair, Mon. Not. R. Astron. Soc. **451**, 2417 (2015).

[6] D. R. Lorimer and M. Kramer, *Handbook of Pulsar Astronomy* (Cambridge University Press, Cambridge, UK, 2004), Vol. 4.

[7] R. W. Hellings and G. S. Downs, Astrophys. J. Lett. **265**, L39 (1983).

[8] H. Xu, S. Chen *et al.* (The CPTA Collaboration), Res. Astron. Astrophys. **23**, 075024 (2023).

[9] J. Antoniadis *et al.* (EPTA and InPTA Collaborations), Astron. Astrophys. **678**, A50 (2023).

[10] G. Agazie *et al.* (The NANOGrav Collaboration), Astrophys. J. Lett. **951**, L8 (2023).

[11] D. J. Reardon, A. Zic *et al.* (The PPTA Collaboration), Astrophys. J. Lett. **951**, L6 (2023).

[12] M. T. Miles *et al.*, Mon. Not. R. Astron. Soc. **536**, 1489 (2024).

[13] G. Agazie *et al.* (The International Pulsar Timing Array Collaboration), arXiv:2309.00693.

[14] J. P. W. Verbiest, S. J. Vigeland, N. K. Porayko, S. Chen, and D. J. Reardon, Results Phys. **61**, 107719 (2024).

[15] T. Damour and A. Vilenkin, Phys. Rev. Lett. **85**, 3761 (2000).

[16] H. Quelquejay Leclere *et al.*, Phys. Rev. D **108**, 123527 (2023).

[17] K. Tomita, Prog. Theor. Phys. **37**, 831 (1967).

[18] S. Matarrese, O. Pantano, and D. Saez, Phys. Rev. D **47**, 1311 (1993).

[19] A. Kosowsky, M. S. Turner, and R. Watkins, Phys. Rev. D **45**, 4514 (1992).

[20] M. Hindmarsh, S. J. Huber, K. Rummukainen, and D. J. Weir, Phys. Rev. Lett. **112**, 041301 (2014).

[21] N. Bartolo, S. Matarrese, A. Riotto, and A. Väihkönen, Phys. Rev. D **76**, 061302 (2007).

[22] L. A. Boyle and A. Buonanno, Phys. Rev. D **78**, 043531 (2008).

[23] L. Sorbo, J. Cosmol. Astropart. Phys. 06 (2011) 003.

[24] J. Antoniadis *et al.* (EPTA and InPTA Collaborations), arXiv:2306.16227.

[25] A. Afzal *et al.* (The NANOGrav Collaboration), Astrophys. J. Lett. **951**, L11 (2023).

[26] Fermi-LAT Collaboration, Science **376**, 521 (2022).

[27] P. Bickel, B. Kleijn, and J. Rice, Astrophys. J. **685**, 384 (2008).

[28] M. Kerr, Astrophys. J. **732**, 38 (2011).

[29] P. Bruel, Astron. Astrophys. **622**, A108 (2019).

[30] M. Kerr, Astrophys. J. **885**, 92 (2019).

[31] M. Kerr, A. Parthasarathy, and T. Cromartie (Large Area Telescope Collaboration), Proc. Sci. ICRC2023 (**2023**) 1595.

[32] J. Luo, S. Ransom, P. Demorest, P. S. Ray, A. Archibald, M. Kerr, R. J. Jennings, M. Bachetti, R. van Haasteren, C. A. Champagne, J. Colen, C. Phillips, J. Zimmerman, K. Stovall, M. T. Lam, and F. A. Jenet, Astrophys. J. **911**, 45 (2021).

[33] A. Susobhanan *et al.*, arXiv:2405.01977.

[34] G. B. Hobbs, R. T. Edwards, and R. N. Manchester, Mon. Not. R. Astron. Soc. **369**, 655 (2006).

[35] R. T. Edwards, G. B. Hobbs, and R. N. Manchester, Mon. Not. R. Astron. Soc. **372**, 1549 (2006).

[36] J. A. Ellis, M. Vallisneri, S. R. Taylor, and P. T. Baker, ENTERPRISE: Enhanced Numerical Toolbox Enabling a Robust PulsaR Inference SuitE, Astrophysics Source Code Library, record ascl:1912.015 (2019), ascl:1912.015.

[37] R. N. Caballero *et al.*, Mon. Not. R. Astron. Soc. **457**, 4421 (2016).

[38] L. Lentati, P. Alexander, M. P. Hobson, S. Taylor, J. Gair, S. T. Balan, and R. van Haasteren, Phys. Rev. D **87**, 104021 (2013).

[39] R. van Haasteren, Y. Levin, P. McDonald, and T. Lu, Mon. Not. R. Astron. Soc. **395**, 1005 (2009).

[40] R. van Haasteren and Y. Levin, Mon. Not. R. Astron. Soc. **428**, 1147 (2012).

[41] R. van Haasteren and M. Vallisneri, Phys. Rev. D **90**, 104012 (2014).

[42] J. Antoniadis *et al.* (EPTA and InPTA Collaborations), 10.5281/zenodo.8164425 (2023).

[43] P. Virtanen *et al.* (SciPy 1.0 Contributors), Nat. Methods **17**, 261 (2020).

[44] R. van Haasteren and M. Vallisneri, Mon. Not. R. Astron. Soc. **446**, 1170 (2015).

[45] B. Allen and J. D. Romano, Phys. Rev. Lett. **134**, 031401 (2025).

[46] M. Crisostomi, R. van Haasteren, P. M. Meyers, and M. Vallisneri, arXiv:2506.13866.

[47] M. Vallisneri and R. van Haasteren, Mon. Not. R. Astron. Soc. **466**, stx069 (2017).
[48] A. E. Hoerl and R. W. Kennard, Technometrics **12**, 55 (1970).
[49] J. Antoniadis *et al.* (EPTA and InPTA Collaborations), Astron. Astrophys. **678**, A48 (2023).
[50] J. Antoniadis *et al.* (EPTA and InPTA Collaborations), Astron. Astrophys. **678**, A49 (2023).
[51] J. Ellis and R. van Haasteren, jellis18/PTMCMCSampler: Official release, 10.5281/zenodo.1037579 (2017).
[52] https://github.com/serevaltolina/PTA_FourierLikelihood
[53] B. Goncharov and S. Sardana, Mon. Not. R. Astron. Soc. **537**, 3470 (2025).
[54] T. Thongmeearkom *et al.*, Mon. Not. R. Astron. Soc. **530**, 4676 (2024).
[55] C. R. Harris *et al.*, Nature (London) **585**, 357 (2020).
[56] J. D. Hunter, Comput. Sci. Eng. **9**, 90 (2007).
[57] D. Foreman-Mackey, J. Open Source Software **1**, 24 (2016).
[58] M. Vallisneri, LIBSTEMPO: Python wrapper for TEMPO2, Astrophysics Source Code Library, record ascl:2002.017 (2020), ascl:2002.017.
[59] J. S. Hazboun, LA_FORGE, 10.5281/zenodo.4152550 (2020).
[60] W. G. Lamb, S. R. Taylor, and R. van Haasteren, Phys. Rev. D **108**, 103019 (2023).
[61] J. W. Cooley and J. W. Tukey, Math. Comput. **19**, 297 (1965).
[62] C. M. Bishop and H. Bishop, *Deep Learning: Foundations and Concepts* (Springer International Publishing, Cham, 2024).
[63] W. Coles, G. Hobbs, D. J. Champion, R. N. Manchester, and J. P. W. Verbiest, Mon. Not. R. Astron. Soc. **418**, 561 (2011).
[64] R. Vershynin, *High-Dimensional Probability: An Introduction with Applications in Data Science*, Cambridge Series in Statistical and Probabilistic Mathematics (Cambridge University Press, Cambridge, England, 2018).
[65] I. Jolliffe and J. Cadima, Phil. Trans. R. Soc. A **374**, 20150202 (2016).